# A novel attention mechanism for noise-adaptive and robust segmentation of microtubules in microscopy images

Achraf Ait Laydi[1,2] [0009-0004-8898-1903], Louis Cueff[1] [0009-0003-6798-0920], Mewen Crespo[1,3] [0000-0003-3834-9462], Yousef El Mourabit[2] [0000-0001-7851-3816] and Hélène Bouvrais[1] [0000-0003-1128-1322]

[1] CNRS, Univ. Rennes, IGDR (Institut de Génétique et Développement de Rennes), UMR6290, 35000 Rennes, France
[2] TIAD Laboratory, Sciences and Technology Faculty, Sultan Moulay Slimane Univ., Morocco.
[3] IRMAR – UMR 6625, 263 avenue du Général Leclerc, 35042, Rennes, France (current affiliation).

## Abstract

**Background:** Segmenting cytoskeletal filaments in microscopy images is essential for studying their roles in cellular processes such as cell division and intracellular transport. However, this task is highly challenging due to the fine, densely packed, and intertwined nature of these structures. Imaging limitations—noise, low contrast, and uneven fluorescence—further complicate analysis. While deep learning has advanced segmentation of large, well-defined biological structures, its performance often degrades under such adverse conditions. Additional challenges include obtaining precise annotations for curvilinear structures and managing severe class imbalance during training.
**Results:** We introduce a novel noise-adaptive attention mechanism that extends the Squeeze-and-Excitation (SE) module to dynamically adjust to varying noise levels. Integrated into a U-Net decoder with residual encoder blocks, this yields ASE_Res_UNet, a lightweight yet high-performance model. To address annotation challenges, we developed a synthetic dataset generation strategy that ensures accurate annotations of fine filaments in noisy images, producing a synthetic dataset with two difficulty levels for segmentation benchmarking. We systematically evaluated loss functions and metrics to mitigate class imbalance, ensuring robust performance assessment. ASE_Res_UNet effectively segmented microtubules in noisy synthetic images, outperforming its ablated variants. It also demonstrated superior segmentation compared to models with alternative attention mechanisms or distinct architectures, while requiring fewer parameters, making it efficient for resource-constrained environments. Evaluation on a newly curated real microscopy dataset and a recently reannotated dataset highlighted ASE_Res_UNet's effectiveness in segmenting microtubules beyond synthetic images. For these datasets, ASE_Res_UNet was competitive with a recent synthetic data-driven approach that shares two cytoskeleton pretrained models. Importantly, ASE_Res_UNet showed strong transferability to other curvilinear structures (blood vessels and nerves) across diverse imaging conditions.
**Conclusions:** This work advances microtubule segmentation through three key contributions: (1) Providing two benchmark datasets (synthetic and real), addressing a critical gap in standardised evaluation resources for this task; (2) Introducing ASE_Res_UNet, a lightweight yet robust model combining noise-adaptive attention with residual learning; (3) Validating competitive performance across synthetic and real microscopy data. Additionally, we demonstrated the robustness and versatility of the proposed architecture across diverse curvilinear segmentation tasks, showcasing potential for broader applications in biological research and medical diagnosis.

## Corresponding author
Hélène Bouvrais: helene.bouvrais@univ-rennes.fr

# 1 Background

Microtubules are essential cytoskeletal filaments involved in a wide range of cellular processes, including tissue modelling, cell division, migration, morphogenesis, and intracellular transport [1-5]. Besides, they serve as therapeutic targets in several diseases, such as cancers and neurodegenerative disorders [6-10], making them a central focus of both fundamental and translational research. Structurally, microtubules are dynamic, semi-flexible filaments composed of tubulin dimers that assemble into a cylindrical structure with a diameter of 25 nm and lengths reaching several tens of micrometres [11]. Their dynamic and mechanical behaviours, such as growth, shrinkage, and bending, are governed by intrinsic properties and regulated by extrinsic factors, including microtubule-associated proteins (MAPs) [12]. To investigate microtubule functions, dynamics, and mechanics in cells, fluorescence microscopy is widely used [13]. In live cells, microtubules can be visualized using fluorescently tagged proteins (e.g., EB proteins to label plus-ends or fluorescent tubulin to label the lattice), while fixed cells are typically stained using anti-tubulin antibodies. Tracking microtubule plus-ends using EB proteins is a well-established method [14, 15] that has enabled detailed characterization of microtubule dynamics and led to the discovery of numerous regulatory MAPs at microtubule extremities [16, 17]. In contrast, the microtubule lattice has received less attention. Recent works, however, have highlighted its regulatory role in processes such as tubulin self-repair and post-translational modifications, which can influence key mechanical properties like flexural rigidity [18-21]. It calls for additional investigations to discover MAPs regulating these processes.

Quantitative analysis of microtubules often requires accurate segmentation to extract geometric and topological features, such as curvature, orientation, and length. However, segmenting microtubules in fluorescence microscopy images remains a major challenge due to several confounding factors. These include high image noise, low contrast, intensity inhomogeneity, and background artifacts. Furthermore, microtubules appear as thin, curved, and sometimes discontinuous filaments, often forming dense, overlapping networks with intensity variation along their length, which complicate their precise delineation. These challenges are exacerbated in live-cell imaging, where low-light acquisition conditions (to avoid phototoxicity) result in low signal-to-noise ratio (SNR) images [22, 23]. Despite the importance of microtubules, relatively few computational tools are available for their full-length segmentation [24]. Existing methods are often multi-step, tailored to specific datasets or applications, require hand-crafted parameter adjustments, and can be difficult to adapt to different microtubule networks or imaging conditions [25-31]. Interestingly, microtubules share structural similarities with other biological curvilinear structures, such as blood vessels, fibrils, or nerve fibres.

Early approaches to curvilinear structure segmentation relied on traditional image processing techniques, particularly filtering methods designed to enhance line-like patterns. One influential method is the vesselness filter introduced by Frangi *et al.*, based on the eigenvalues of the Hessian matrix [32]. Later enhancements included orientation- and symmetry-based filters, matched filters, and multi-scale Gabor filters [33-37]. Active contours and geodesic path-based algorithms have also been applied to delineate elongated structures by leveraging local continuity and smoothness [29, 38-42]. For cytoskeleton segmentation, several traditional methods have been developed, each having some limitations. FIESTA fits 2D Gaussian distributions to image intensities but is restricted to sparse networks and requires 21 user-defined parameters [43]. SIFNE employs linear and oriented filter transforms followed by a fragment reconstruction step based on geometric constraints that is sensitive to noisy backgrounds [28]. TSOAX uses stretching open active contours but requires users to define 26 parameters, limiting its adaptability [42]. FinTA applies vectorial tracing for improved spatial precision but depends on eight predefined parameters [44]. ILEE uses adaptive local thresholding to handle high dynamic ranges in filament intensities but struggles with high noise levels [45]. Filament Sensor 2.0 enables single-filament frame-to-frame tracking but may not perform well with dense or highly curved filaments [46]. Overall, these methods depend heavily on hand-crafted parameters, are noise sensitive, and often perform poorly on diverse datasets due to limited robustness and generalisability. Indeed, they lack the contextual awareness that can be captured by deep learning approaches.

The rise of deep learning, particularly convolutional neural networks (CNNs), has significantly advanced biomedical and microscopy image segmentation. The introduction of U-Net by Ronneberger *et al.* was a breakthrough, enabling accurate, end-to-end segmentation with limited annotated data [47, 48]. Its encoder-decoder architecture with skip connections facilitates precise localisation while preserving contextual information. Numerous U-Net variants have since been proposed, including models adapted to curvilinear structure segmentation [49]. For instance, Residual U-Net introduces residual connections to capture long-range dependencies [50, 51], while Attention U-Net incorporates gating mechanisms to focus on informative features [52-54]. However, these deep learning advances have had only limited impact on the analysis of cytoskeletal filaments in fluorescence microscopy [31, 55-57], especially compared to their widespread application in other biomedical domains. The two first deep-learning based tools for segmenting cytoskeletal filaments are those from Liu *et al*. in 2018 [58] and Masoudi *et al.* in 2019 [59]. Liu and co-workers implemented a modified U-Net to segment microtubules and actin filaments. While this approach outperformed traditional methods like SOAX, it was only tested on low-noise images, limiting its real-world applicability. Masoudi *et al.* used a recurrent neural network to track microtubules in *in vitro* images, where filaments were sparse and clearly visible. However, its performance on dense *in cellulo* cytoskeletal networks remains unexplored. Very recently, FAST, which employs a UNet++ model, enabled the segmentation and classification of actin structures [60]. However, it is highly specific to this application. A persistent challenge in these deep learning approaches is their reliance on large, manually annotated datasets, which are time-consuming to generate and dependent on expert input. To address this, SynSeg introduced a framework that generates synthetic datasets to train a U-Net architecture [57]. While SynSeg provided pre-trained models that performed well on the Higaki dataset of microtubules, this dataset lacks fluorescence background noise, making it a less challenging benchmark for real-world applications.

Overall, several key challenges have to be addressed for segmenting microtubules or other curvilinear structures in noisy and low-contrasted images. (i) High-quality annotated datasets are scarce due to the difficulty of labelling fine structures, particularly in live-cell images. (ii) Severe class imbalance, where curvilinear structures occupy only a small fraction of the image, complicates model training. (iii) Many existing models are modality-specific and do not generalise well across imaging conditions. (iv) Incorporating residual or attention mechanisms improves model performance but increases parameter count and computational demand, limiting usability in resource-constrained environments. (v) Segmenting faint, noisy curvilinear structures in low-SNR fluorescence images remains inherently difficult.

To address these challenges, we presented the following contributions. (1) We propose a novel noise-adaptive attention mechanism to improve segmentation of thin structures in noisy images. (2) We introduce ASE_Res_UNet, a deep-learning architecture based on U-Net with residual blocks in the encoder and Adaptive Squeeze-and-Excitation (ASE) attention blocks in the decoder. (3) We generated original synthetic datasets of fluorescence microscopy images of microtubules (*MicSim_FluoMT*) with two difficulty levels and high-quality ground truth masks that are rarely available in real noisy microscopy images. ASE_Res_UNet accurately segmented microtubules in these synthetic images and it outperformed standard U-Net, its own architectural variants, and several architectures based on alternative attention mechanisms. (4) We demonstrated strong transferability of ASE_Res_UNet across diverse curvilinear structures and imaging modalities, including fluorescent microtubule images (providing a novel dataset, *MicReal_FluoMT*), retinal blood vessels (*DRIVE* dataset) [61], and corneal nerves (*CORN-1* dataset) [62]. (5) We tackled the class imbalance problem by selecting suitable loss function and evaluation metrics, avoiding misleading performance estimates due to dominant background class. Overall, this work makes a significant contribution to the field of computational bioimage analysis by providing ASE_Res_UNet—a robust and versatile deep learning framework designed for the accurate segmentation of curvilinear structures. It not only enhances the quantitative analysis of cytoskeletal filaments, thereby advancing our fundamental understanding of various cellular processes, but also holds promise for accelerating disease diagnosis by enabling precise delineation of diagnostically relevant structures in biomedical images.

# 2 Methods

## 2.1 Synthetic datasets for segmenting microtubules

To address the challenges of microtubule segmentation in noisy fluorescence microscopy images, we created two synthetic datasets of fluorescently labelled microtubules, along with corresponding binary masks (ground truth) using a two-step pipeline (details in Supplemental Material 1A). (1) *Cytosim* was used to mimic the astral microtubule network of *Caenorhabditis elegans* zygotes (Fig. S1A) [63]. (2) *ConfocalGN* converted the simulations into realistic confocal-like fluorescent images (Fig. S1A-E) [64]. This pipeline yielded two large, fully annotated datasets with distinct fluorescence intensity distributions along cytoskeletal filaments, leading to different difficulty levels for segmentation (Fig. S2) [65]. In the *MicSim_FluoMT easy* dataset, microtubules display uniform intensity along their length (Fig. S2A, D). In the *MicSim_FluoMT complex* dataset, intensity decreases toward the cell periphery, making microtubule extremities harder to distinguish from the cytoplasm background (Fig. S2B, E). Each dataset contains 1192 images with diversity in microtubule density, curvature, and crossing patterns. We used 953 images for training, 119 for validation, and 120 for testing. We also benchmarked our architecture on the recently published SynthMT dataset, which comprises 6600 synthetic interference reflection microscopy (IRM) images containing microtubules [31]. The dataset was split as follows: 5280 images (80%) for training, 660 (10%) for validation, and 660 (10%) for testing.

## 2.2 Real image datasets for testing model generalisation

To evaluate performance on real microscopy data, we created the *MicReal_FluoMT* dataset consisting of 49 fluorescence microscopy images of *C. elegans* zygotes, where microtubules were stained with anti-tubulin antibodies. Annotation was performed using a previously developed three-step pipeline [66] (details in Supplemental Material 1B). The dataset was split into 19 training, 10 validation, and 10 test images [67]. Some masks included annotations that were distinct from microtubules, due to non-specific staining or reflecting the annotation-pipeline limitations. We also used in-house image samples to test the model's robustness, namely live *C. elegans* zygotes expressing GFB::TBB-2 imaged using the Nikon Nsparc microscope. Finally, we benchmarked our architecture on an independent microtubule dataset, the Higaki 2024 dataset [56], whose test split was recently reannotated to higher mask accuracy [57]. This update included 98 confocal images of *Nicotiana tabacum* BY-2 cells labelled with YFP-tubulin. We used 90 for 5-fold cross-validation and reserved 8 for segmentation predictions.

To further assess the model's transferability beyond microtubules, we tested its performance on two public datasets, selected for their structural similarities to cytoskeletal filaments. The *DRIVE* (Digital Retinal Images for Vessel Extraction) dataset comprises 40 colour fundus images of retinal vessels [61], widely used in vessel segmentation benchmarks [68-70]. Challenges include low-contrast, illumination variations, complex vessel branching, and diverse vessel widths. We trained the model on 20 images, and we used 10 images for validation and 10 for testing. The *CORN-1* dataset contains 1516 confocal microscopy images of the corneal nerves with manual annotations [62] and was used in a few deep-learning based segmentation studies [68, 71]. These images pose challenges due to noise, background inhomogeneity and variability in nerve thickness and intensity. Training was performed using a 5-fold cross-validation approach.

## 2.3 ASE_Res_UNet architecture including a novel noise-adaptive attention mechanism.

We developed ASE_Res_UNet (Fig. 1), an U-Net variant whose originality lies in the use of a novel attention module–Adaptive Squeeze-and-Excitation (ASE)–which enables adaptation to noise [72, 73]. The placement of modules within the U-Net backbone was carefully designed to maintain a reasonable number of trainable parameters.

The encoding path consists of four down-sampling blocks, starting with 32 filters and doubling them at each stage. Each block comprises two convolutional layers, each followed by batch normalization and ReLU activation, and a residual block that incorporates a skip connection [74] (Eq. 1). The later preserves low-level spatial information and facilitates gradient flow during training, which was crucial for learning fine, elongated patterns in noisy environments. By incorporating residual units early in the network, we ensured that critical filament features were retained despite successive down sampling operations.

$$y = \mathrm{ReLU}\left(\mathrm{BN}\left(W_2 * \mathrm{ReLU}\big(\mathrm{BN}(W_1 * x)\big)\right) + x\right) \qquad (1)$$

where *x* denotes the input, *W1* and *W2* are the weights of the first and second convolutional layers, respectively, and BN refers to Batch Normalisation.

The decoding path consists of four up-sampling blocks, mirroring the encoder, which includes transposed convolution layers for up-sampling, concatenation with the corresponding encoder feature map, the Adaptive Squeeze-and-Excitation (ASE) attention module, and two convolutional layers with batch normalisation and ReLU activation (Fig. 1). We modified the traditional Squeeze-and-Excitation attention mechanism, which statically recalibrates channel-wise responses [72], into an adaptive version that dynamically adjusts attention weights based on the input's noise characteristics, making the decoding process context-sensitive.

In more detail, the noise level *N* is estimated from the input using a convolutional layer followed by a sigmoid activation (Eq. 2). This estimate correlates with the actual image noise, as demonstrated by the values computed for the example images in Figure S2 (A1: 0.336, B1: 0.346, A2: 0.888, B2: 0.882). This alignment confirmed that *N* provides a meaningful representation of noise levels in the input images, supporting its use in the adaptive noise estimation framework of ASE_Res_UNet. Then, the noise level is averaged over spatial dimensions and batch (Eq. 3). The reduction ratio $r_{adaptive}$ is dynamically adjusted based on the averaged noise level, using the same reduction ratio of 16 as in [72], since this value was shown to achieve a good trade-off between accuracy and model complexity (Eq. 4). A global spatial average pooling computes the channel-wise statistics (Eq. 5). Two fully connected layers with a ReLU and sigmoid activation compute the channel-wise weights $e_c$ (Eq. 6). Finally, the input *X* is rescaled using $e_c$ (Eq. 7).

$$N = Sigmoid\big(Conv2d(X)\big), N \in R^{B\times H\times W} \qquad (2)$$

$$\widehat{N} = \frac{1}{BxHxW}\sum_{b=1}^{B}\sum_{i=1}^{H}\sum_{j=1}^{W} N_{b,i,j} \qquad (3)$$

$$r_{adaptive} = max\big(1, \lceil 16 \cdot \big(1 + \widehat{N}\big)\rceil\big) \qquad (4)$$

$$s_c = \frac{1}{H \times W}\sum_{i=1}^{H}\sum_{j=1}^{W} X_{c,i,j}\,, \forall c \in \{1, \dots, C\} \quad (5)$$

$$e_c = Sigmoid\left(FC_2^{(r)} \cdot ReLU\left(FC_1^{(r_{adaptive})} \cdot s_c\right)\right) \qquad (6)$$

$$\widetilde{X_{c,\iota,\jmath}} = X_{c,i,j} \cdot e_c, \forall c, i, j. \qquad (7)$$

ASE modules were applied only in the decoder, where spatial details are reconstructed. This design should promote continuity of thin structures while suppressing background noise during up sampling.

To our knowledge, ASE_Res_UNet is the first architecture to incorporate noise-adaptive channel attention into a U-Net -based model specifically tailored for curvilinear structure segmentation.

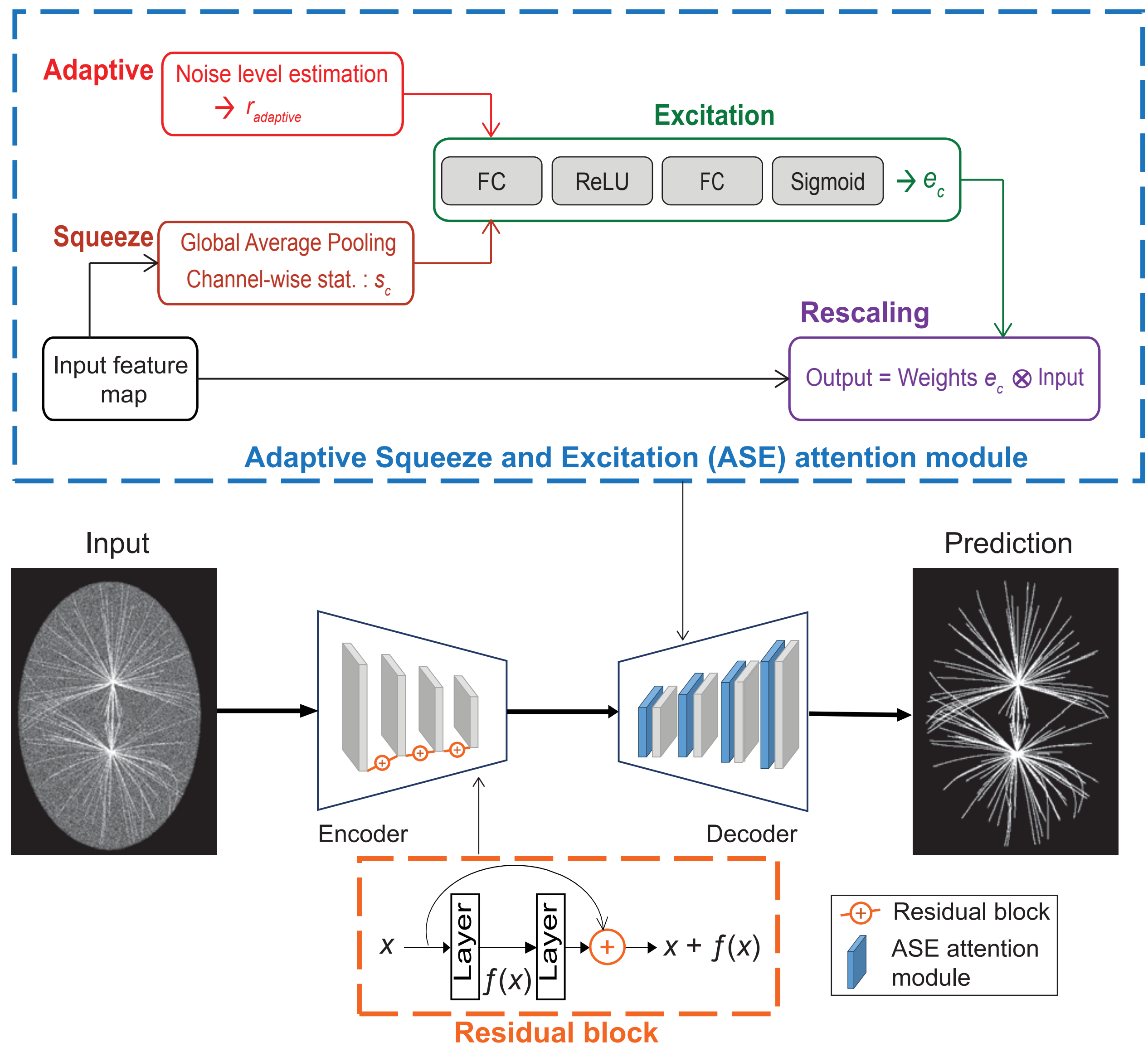

**Fig. 1: ASE_Res_UNet includes noise-adaptive attention mechanisms in the decoder.** Proposed architecture of ASE_Res_UNet, including (orange) the residual blocks in the encoder and (blue) the novel Adaptive Squeeze-and-Excitation (ASE) attention modules in the decoder. Each ASE module includes: (brown) a squeeze step which uses global average pooling per channel to compute channel-wise statistics $s_c$, and (red) an adaptive step, which assesses the noise level to determine the reduction ratio $r_{adaptive}$. This enables (green) the computation of the channel-wise weights $e_c$, which are used (purple) to rescale the input map.

### 2.4 Handling background class dominance in metric and loss function selection

To assess the segmentation performance, we employed six quantitative metrics (Table S1), each capturing different aspects of prediction quality: Dice Similarity Coefficient, Intersection over Union (IoU), Sensitivity, Precision, Matthews Correlation Coefficient (MCC), and Area Under the Precision-Recall Curve (PR AUC) [75, 76]. These metrics were appropriate for highly imbalanced datasets [77, 78]. By combining both threshold-dependent and threshold-independent metrics, we ensured a more robust and holistic performance assessment. Statistical significance was evaluated using two-tailed Student's *t*-test with Welch–Satterthwaite correction for unequal variance. We reported significance as follows: ◊: $0.01 < p \leq 0.05$; *: $0.001 < p \leq 0.01$; **: $1 \times 10^{-4} < p \leq 0.001$; ***: $p \leq 1 \times 10^{-4}$; and ns (non-significant): $p > 0.05$.

To handle class imbalance, we evaluated five loss functions on the *MicSim_FluoMT complex* dataset, under identical training conditions (section 2.5): Binary Cross Entropy (BCE), Dice Loss, Focal Loss, Hausdorff Distance Loss, and Weighted Cross Entropy (WCE) [79, 80]. WCE yielded the best overall performance (Table 1).

Foreground-to-background weights were empirically tuned, with 1.0 for microtubules and 0.25 for background producing the best results.

| Model | Dice ↑ | IoU ↑ | Sensitivity ↑ | Precision ↑ | MCC ↑ | PR AUC ↑ |
|---|---|---|---|---|---|---|
| WCE | **0.7825 ± 0.0156** | **0.9688 ± 0.0011** | **0.7345 ± 0.0233** | 0.8377 ± 0.0108 | **0.7762 ± 0.0148** | **0.8730 ± 0.0147** |
| BCE | 0.7625 ± 0.0169<br>*** | 0.9687 ± 0.0010 | 0.6589 ± 0.0221<br>*** | 0.9051 ± 0.0090<br>*** | 0.7647 ± 0.0154<br>*** | 0.7764 ± 0.0271<br>*** |
| Dice | 0.6898 ± 0.0340<br>*** | 0.9607 ± 0.0033<br>*** | 0.5775 ± 0.0607<br>*** | 0.8666 ± 0.0357<br>*** | 0.6966 ± 0.0261<br>*** | 0.7432 ± 0.0216<br>*** |
| Focal | 0.7196 ± 0.0173<br>*** | 0.9654 ± 0.0012<br>*** | 0.5814 ± 0.0209<br>*** | **0.9447 ± 0.0071**<br>*** | 0.7336 ± 0.0149<br>*** | 0.8308 ± 0.0152<br>*** |
| Hausdorff Distance | 0.7422 ± 0.0166<br>*** | 0.9611 ± 0.0015<br>*** | 0.7354 ± 0.0195 | 0.7494 ± 0.0196<br>*** | 0.7320 ± 0.0166<br>*** | 0.8193 ± 0.0187<br>*** |

**Table 1: The Weighted Cross Entropy loss outperforms the other loss functions on segmenting microtubules.**
Microtubule segmentation performance of ASE_Res_UNet under different loss functions including the Weighted Cross Entropy (WCE) loss, the Binary Cross Entropy (BCE) loss, the Dice loss, the Focal loss and the Hausdorff loss, evaluated on the *MicSim_FluoMT complex* dataset using various metrics (mean ± standard deviation over 120 test images). Bold values indicate the best performances. Statistical differences between WCE model and other models using different loss functions are indicated only when significant (***: $p \leq 0.0001$).

### 2.5 Training

During training, we tuned several hyperparameters to balance effective multi-scale feature extraction with moderate computational cost. The best configuration included: an initial learning rate of 0.001, dynamically reduced based on validation performance; a batch size of 2, balancing memory efficiency and convergence speed; the Adam optimizer configured with $\beta_1$ and $\beta_2$ equal to 0.9 and 0.999; and a 3x3 filter size, chosen for its low sensitivity to noise and effectiveness in capturing fine details. We applied data augmentation during training, including random rotations, horizontal and vertical flips, and slight colour jittering. Training was terminated when the validation loss stabilised over 50 consecutive epochs, indicating convergence. This resulted in a model with approximately 8 million trainable parameters, more compact than models such as TransUNet or CAR-UNet models [81, 82].

## 3 Experimental results

### 3.1 ASE_Res_UNet effectively segments microtubules from synthetic microscopy images mimicking the microtubule networks of *C. elegans* zygotes.

Various U-Net variants incorporating residual and attention modules have been developed or used for segmenting retinal blood vessels [49, 50, 52-54, 68, 69, 83-86]. We hypothesised that a similar approach could be effective for segmenting microtubules, given geometrical similarities both being tiny and curvilinear structures. To design a U-Net-based architecture tailored for microtubule segmentation in noisy fluorescence microscopy images, we trained on synthetic image datasets with reliable ground-truth annotations, allowing robust performance evaluation (cf. Section 2.1 and Supplementary Material 1A). We developed a novel model, ASE_Res_UNet, which combines residual encoder blocks with ASE attention decoder modules—a newly proposed attention mechanism designed to account for noise variability in microscopy images (cf. section 2.3, Fig. 1).

We first asked whether ASE_Res_UNet would be effective in segmenting microtubules in fluorescence microscopy images using *MicSim_FluoMT easy* dataset. This dataset simulates cytoplasmic fluorescence background, such as that arising from fluorescently tagged tubulin (e.g., GFP::tubulin), which can interfere with accurate microtubule segmentation. The model was trained using the WCE loss function (cf. sections 2.4 & 2.5) with no image preprocessing—training conditions that were maintained across all models in this study. ASE_Res_UNet successfully segmented microtubules along their full length (Fig. 2B). Quantitative evaluation confirmed strong performance: Dice = 0.9308 $\pm$ 0.0066, IoU = 0.9894 $\pm$ 0.0005, sensitivity = 0.9295 $\pm$ 0.0070, and precision = 0.9322 $\pm$ 0.0077 (mean $\pm$ standard deviation (SD) across 120 test images, cf. section 2.4). These results indicated a well-balanced segmentation, with neither false positives nor false negatives being predominant. Given the class imbalance in our images, we also computed the Matthews Correlation Coefficient (MCC = 0.9281 ± 0.0067) and the area under the Precision-Recall curve (PR AUC = 0.9845 ± 0.0029), both of which further supported the model's excellent performance. Importantly, segmentation quality remained high even under challenging conditions rarely addressed in previous work on blood vessel segmentation, such as high filament density in central regions and filament crossings in periphery. These results demonstrated that ASE_Res_UNet is a robust and accurate model for segmenting microtubule networks in fluorescence microscopy images, even in the presence of cytoplasmic background fluorescence.

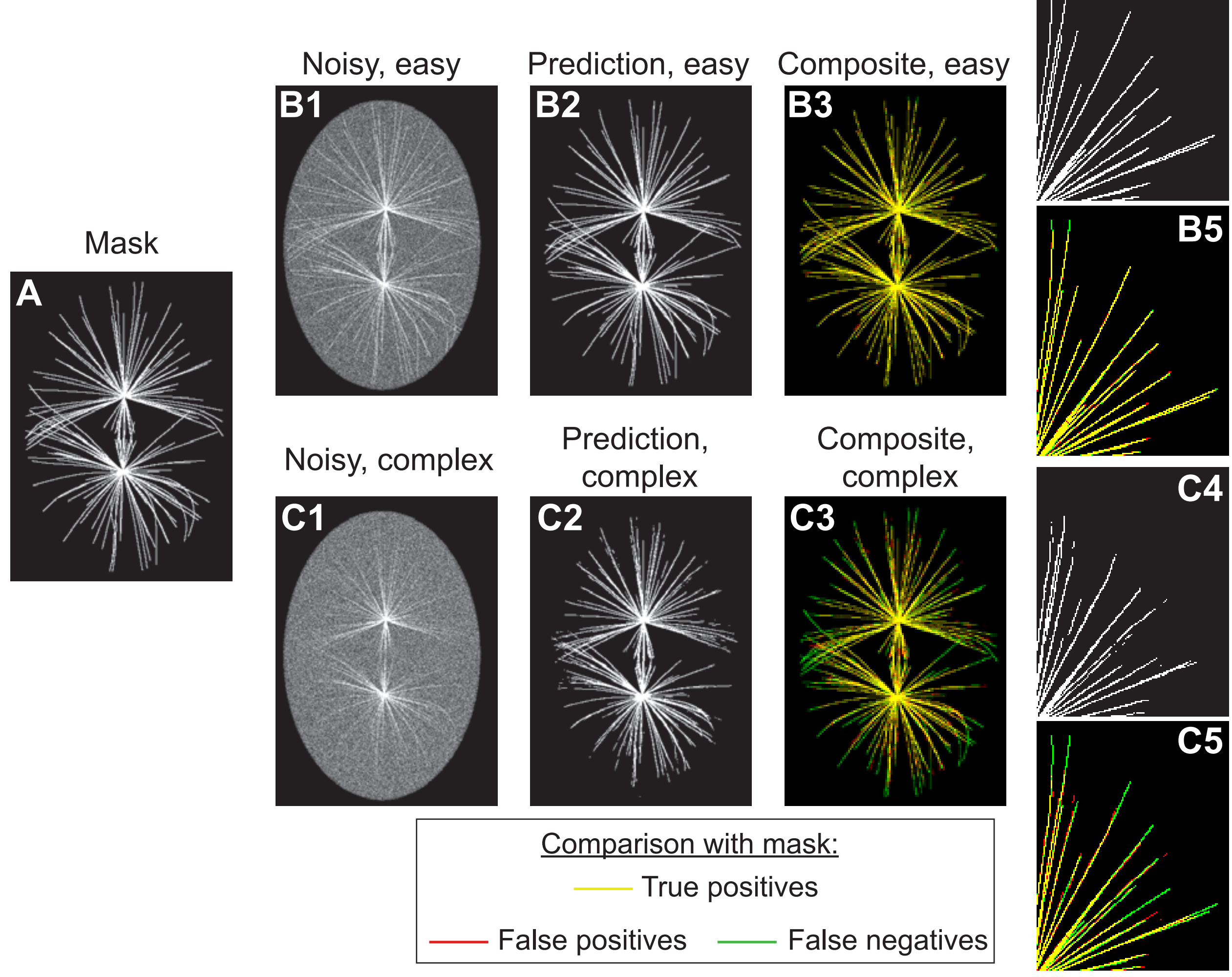

**Fig. 2: ASE_Res_UNet architecture accurately segments microtubules in noisy fluorescence images, though performance declines in regions with extremely low signal-to-noise ratios.**

Microtubule segmentation results obtained using ASE_Res_UNet on a test image from the *MicSim_FluoMT* datasets: (**A**) ground truth; (**B1, C1**) *easy* and *complex* images; (**B2, C2**) corresponding predicted images; (**B3, C3**) composite images used to visualise segmentation accuracy: true positives are shown in yellow, false negatives in green, false positives in red, and true negatives in black; (**B4, C4**) zoomed-in regions of interest (ROI); (**B5, C5**) composite ROI images.

To further assess the model under more challenging conditions, we created a second synthetic dataset, *MicSim_FluoMT Complex*, which simulates decreasing fluorescence intensity toward microtubule ends. This mimics the limited visibility of peripheral microtubules in confocal microscopy, caused by focal plane limitations and microtubule bending. These images contain regions where microtubules are visually indistinguishable from cytoplasmic background, presenting extremely low SNR scenario. These variations in signal intensities were of particular interest to rigorously test the noise-adaptability of ASE_Res_UNet. The model was trained on this dataset and then evaluated on the corresponding test images. While segmentation remained accurate in the central regions of embryos, errors were frequently observed at microtubule extremities (Fig. 2C). Performance metrics reflected the increased difficulty: Dice = 0.7825 ± 0.0156, IoU = 0.9688 ± 0.0011, sensitivity = 0.7345 ± 0.0233, precision = 0.8377 ± 0.0108, MCC = 0.7762 ± 0.0148, and PR AUC = 0.8730 ± 0.0147 (mean ± SD over 120 test images). Interestingly, ASE_Res_UNet tended to limit false positives, as indicated by higher precision relative to sensitivity. This conservative segmentation behaviour was desirable in biological applications, where microtubules are typically the primary structures of interest and false positives could be misleading.

We therefore concluded that the proposed ASE_Res_UNet architecture was well-suited for accurately segmenting microtubules in noisy fluorescence images. However, its performance declined in extreme cases where microtubules were barely distinguishable from the fluorescence background. These findings motivated us to further investigate the contribution of each architectural component of ASE_Res_UNet to its overall segmentation performance.

### 3.2 Both residual blocks and ASE modules enhance segmentation of low-intensity microtubule extremities.

| Model | Loss ↓ | Dice ↑ | IoU ↑ | Sensitivity ↑ | Precision ↑ | MCC ↑ | PR AUC ↑ |
|---|---|---|---|---|---|---|---|
| U-Net | **0.0035 ± 0.0002** | **0.9315 ± 0.0060** | 0.9894 ± 0.0004 | **0.9343 ± 0.0084** *** | 0.9289 ± 0.0088 * | **0.9288 ± 0.0060** | **0.9849 ± 0.0028** |
| Res_UNet | 0.0038 ± 0.0003 | 0.9287 ± 0.0071 ◊ | 0.9891 ± 0.0005 *** | 0.9245 ± 0.0100 *** | **0.9329 ± 0.0084** | 0.9258 ± 0.0071 ◊ | 0.9835 ± 0.0035 |
| ASE_UNet | 0.0036 ± 0.0002 | 0.9314 ± 0.0059 | 0.9894 ± 0.0004 | 0.9317 ± 0.0065 * | 0.9311 ± 0.0066 | 0.9286 ± 0.0059 | 0.9843 ± 0.0029 |
| ASE_Res_UNet | 0.0036 ± 0.0002 | 0.9308 ± 0.0066 | 0.9894 ± 0.0005 | 0.9295 ± 0.0070 | 0.9322 ± 0.0077 | 0.9281 ± 0.0067 | 0.9845 ± 0.0029 |

**Table 2: All ASE_Res_UNet variants accurately segment microtubules in the *MicSim_FluoMT easy* dataset.**
Microtubule segmentation performance of ASE_Res_UNet and its variant architectures on the *MicSim_FluoMT easy* dataset, evaluated using various metrics (mean ± standard deviation over 120 test images). Bold values indicate the best performances. Statistical differences between ASE_Res_UNet and its variant architectures are indicated only when significant (◊: $0.01 < p \leq 0.05$; *: $0.001 < p \leq 0.01$; ***: $p \leq 0.0001$).

To evaluate the contribution of each component in ASE_Res_UNet, we conducted an ablation study by designing three variant architectures for comparison: U-Net, the baseline architecture; Res_UNet, incorporating residual blocks in the encoder, mirroring ASE_Res_UNet's encoder path; and ASE_UNet, integrating ASE attention modules in the decoder, mimicking ASE_Res_UNet's decoder path. We first applied this ablation study to the *MicSim_FluoMT easy* dataset. The baseline U-Net already achieved high performance across all evaluation metrics (Table 2), suggesting that the U-Net backbone, the loss function, and the training setup were well optimised. Adding residual blocks (Res_UNet), ASE attention modules (ASE_UNet), or both (ASE_Res_UNet) did not significantly improved performance. Visual inspection of the predictions confirmed these results (Fig. S3), indicating that this relatively easy dataset did not sufficiently challenge the architectures. In this case, the standard U-Net was capable of capturing most microtubule structures, leaving little room for further gains.

| **Model** | **Loss ↓** | **Dice ↑** | **IoU ↑** | **Sensitivity ↑** | **Precision ↑** | **MCC ↑** | **PR AUC ↑** |
|---|---|---|---|---|---|---|---|
| U-Net | 0.0136 ± 0.0011 *** | 0.7362 ± 0.0190 *** | 0.9662 ± 0.0022 *** | 0.6175 ± 0.0293 *** | **0.9132 ± 0.0147 ***** | 0.7429 ± 0.0158 *** | 0.8657 ± 0.0140 ** |
| Res_UNet | 0.0132 ± 0.0007 *** | 0.7383 ± 0.0185 *** | 0.9660 ± 0.0016 *** | 0.6292± 0.0271 *** | 0.8943 ± 0.0127 *** | 0.7419 ± 0.0160 *** | 0.8557 ± 0.0162 *** |
| ASE_UNet | 0.0125 ± 0.0007 *** | 0.7694 ± 0.0155 *** | 0.9679 ± 0.0014 *** | 0.7025 ± 0.0369 *** | 0.8544 ± 0.0383 *** | 0.7658 ± 0.0142 *** | 0.8635 ± 0.0157 *** |
| ASE_Res_UNet | **0.0116 ± 0.0004** | **0.7825 ± 0.0156** | **0.9688 ± 0.0011** | **0.7345 ± 0.0233** | 0.8377 ± 0.0108 | **0.7762 ± 0.0148** | **0.8730 ± 0.0147** |

**Table 3: ASE_Res_UNet outperforms its architectural variants in microtubule segmentation on the *MicSim_FluoMT complex* dataset.**
Microtubule segmentation performances of ASE_Res_UNet and its variant architectures on the *MicSim_FluoMT complex* dataset, evaluated using various metrics (mean ± standard deviation over 120 test images). Bold values indicate the best performances. Statistical differences between ASE_Res_UNet and its variant architectures are indicated only when significant (**: $0.0001 < p \leq 0.001$; ***: $p \leq 0.0001$).

We hypothesised that the more challenging *MicSim_FluoMT Complex* dataset—with variable fluorescence intensity along microtubules—would better reveal performance differences. Indeed, ASE_Res_UNet outperformed the baseline U-Net across all metrics, except for precision (Table 3). This indicated that combining residual blocks and ASE modules enhanced the model's ability to segment microtubules in noisy and variable-intensity conditions. We interpreted the slightly lower precision as a trade-off: reducing false negatives came at the cost of a slight increase in false positives. When only residual blocs were added to the encoder (Res_UNet), we observed a modest performance improvement over U-Net (Table 3). Adding only ASE attention modules into the decoder (ASE_UNet) resulted in a more substantial boost, highlighting the importance of adaptive attention for detecting faint or low-contrast filaments (Table 3). Notably, ASE_Res_UNet outperformed ASE_UNet, confirming that residual connections and attention mechanisms provided complementary benefits for robust segmentation under complex scenarios. To complement the quantitative analysis, we visually examined the segmentation outputs across several test images. ASE_Res_UNet showed better recovery of microtubule extremities compared to U-Net (Fig. 3C1-C3), consistent with confusion matrix results (Table S3). U-Net missed microtubule pixels in filament regions with lower intensity near the periphery. To explore this spatial effect further, we computed performance metrics separately for two embryo regions: the periphery, where microtubules extremities are faint, and the central region, where microtubules are more prominent (Fig. S4). For ASE_Res_UNet, segmentation quality was consistently lower in peripheral regions compared to central ones—as reflected by reduced Dice, sensitivity, MCC, and PR AUC scores (Table S4). When comparing ASE_Res_UNet to U-Net across these two regions, the performance gain was more pronounced in the periphery, suggesting that the added modules particularly benefited segmentation of low-intensity microtubule ends. We next examined whether the individual components—residual blocks or ASE modules—were sufficient to improve peripheral segmentation. Visually, Res_UNet underperformed compared to ASE_Res_UNet (Fig. 3D1–D3). Although it slightly improved segmentation of some extremities over U-Net, it failed to generalise across all noisy regions. This was supported by the spatial analysis, which showed only minor improvements over U-Net in the periphery (Table S4). In contrast, ASE_UNet showed stronger enhancement of extremity segmentation (Fig. 3E1–E2), albeit with an increase in false positives (Table S3). Spatial metrics confirmed that ASE modules significantly improved performance in the periphery (Table S4). Still, ASE_UNet underperformed relative to ASE_Res_UNet (Fig. 3E vs. 3F), suggesting that combining residual and attention modules yielded better results than either component alone.

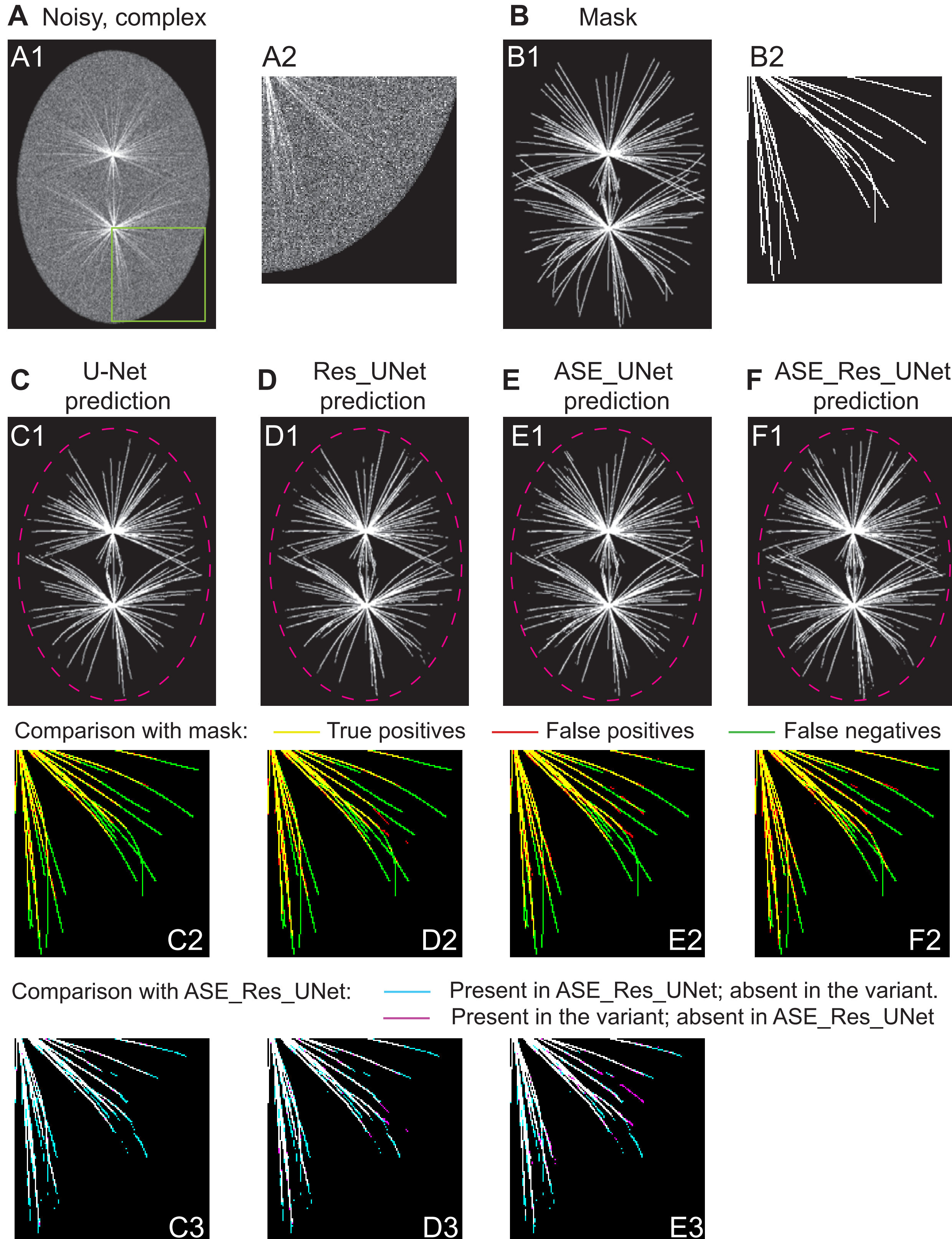

**Fig. 3: ASE_Res_UNet recovers longer microtubules of the *MicSim_FluoMT complex* dataset compared to its variants.**
Microtubule segmentation results obtained using ASE_Res_UNet and its variants on a sample test image from the *MicSim_FluoMT complex* dataset. (**A**) Input image; (**B**) Corresponding ground truth; (**C-F**) Predicted segmentations from (C) U-Net, (D) Res_UNet, (E) ASE_UNet, and (F) ASE_Res_UNet models. (A1-F1): whole simulated embryo; (A2-F2): zoomed-in regions of interest (ROI) to better highlight differences between model predictions and ground truth, with cropping window shown in green on panel A1; (C2-F2) composite ROI images showing true positives in yellow, false negatives in green, false positives in red, and true negatives in black; (C3-E3): composite ROI comparison between (magenta) each variant and (cyan) ASE_Res_UNet. Therefore, microtubules segmented by both architectures appear in white, those segmented only by the variant architecture appear in magenta, and those segmented only by ASE_Res_UNet appear in cyan. Dashed pink line indicates the embryo contour.

Overall, both quantitative and qualitative analyses demonstrated that ASE_Res_UNet improved segmentation accuracy and robustness to noise, particularly for faint microtubule extremities. We proposed that the combination of residual blocks and ASE attention modules enhanced feature extraction and noise handling, allowing the model to better distinguish microtubules from background fluorescence. Since our results highlighted persistent challenges in highly degraded regions, where microtubule signals are barely detectable, we questioned whether alternative model architectures might offer improved performance under these extreme conditions.

### 3.3 ASE_Res_UNet outperforms other advanced architectures in segmenting low-intensity microtubule extremities.

Deep learning models have been developed for curvilinear structure segmentation such as retinal blood vessels—morphologically similar to microtubules—and have thus inspired a variety of architectural approaches from which we could draw [52, 68, 69, 71, 82, 83, 87-89]. We benchmarked ASE_Res_UNet against several deep-learning architectures that differ in the attention mechanisms or in their core component. We used the *MicSim_FluoMT complex* dataset, which represents a highly challenging segmentation scenario. All models were trained under identical conditions (sections 2.4 & 2.5) and evaluated using both quantitative metrics and qualitative visual inspection.

We first focussed on critically evaluating key architectural design choices: (1) our custom attention mechanism, and (2) the placement of residual and attention blocks within the network. To do this, we compared ASE_Res_UNet against three deep learning architectures that also employed attention mechanisms and residual components within an U-Net backbone: CAR-UNet [82], AG_Res_UNet [90], and SE_Res_UNet [72]. CAR-UNet (Channel Attention Residual U-Net), originally developed for segmenting retinal vessels with strong performance [82], integrates Channel Attention Double Residual Blocks (CADRB) in both the encoder and decoder paths, and Modified Efficient Channel Attention (MECA) modules in the skip connections (Fig. S5A). In contrast to ASE_Res_UNet, CAR-UNet applies both residual and attention components at every layer, resulting in a larger model (~16 million parameters). AG_Res_UNet incorporates a grid-based self-attention gating module, previously applied in biomedical segmentation tasks [90]. In this comparison, we retained the same backbone as ASE_Res_UNet, and modified only the attention mechanism to focus on our novel mechanism. This resulted in a model with a similar number of parameters (Fig. S5B). Similarly, SE_Res_UNet was created by replacing the ASE module with a Squeeze-and-Excitation (SE) module [72] (Fig. S5C), allowing us to isolate the contribution of our adaptive SE module. SE mechanisms have also been applied in retinal vessel segmentation [83, 84]. Among these three models, ASE_Res_UNet achieved the best performance across multiple metrics, including Dice coefficient, IoU, sensitivity, MCC, and PR AUC (Table 4). The most notable gain was in sensitivity, indicating improved detection of low-intensity microtubule extremities. This came with a minor reduction in precision due to increased false positives (Table S5), yet overall performance was superior without increasing model size. Qualitative analysis of segmentation predictions further supported these findings. CAR-UNet and AG_Res_UNet both produced shorter segmented filaments (Fig. 4D, E), reflecting their lower sensitivity and higher false negative counts. In contrast, SE_Res_UNet produced longer filaments (Figure 4F), demonstrating the utility of SE blocks for enhancing low-intensity detection. Still, it lagged behind ASE_Res_UNet overall.

Altogether, these comparisons validated two key design choices in ASE_Res_UNet: (1) ASE modules, which integrate noise estimation, outperformed AG and SE modules in noisy images. (2) Placing residual blocs in the encoder and attention mechanisms in the decoder proved sufficient, enabling high performances with a compact model suitable for resource-constrained environments. Importantly, ASE_Res_UNet achieved the best performance in terms of training and inference speed.

| **Model** | **Parameter number ↓** | **Loss ↓** | **Dice ↑** | **IoU ↑** | **Sensitivity ↑** | **Precision ↑** | **MCC ↑** | **PR AUC ↑** | **Training time ↓** | **Inference rate ↑** (image/s) |
|---|---|---|---|---|---|---|---|---|---|---|
| CAR-UNet | 15.83 millions | 0.0127 ± 0.0005 *** | 0.7605 ± 0.0147 *** | 0.9676 ± 0.0013 *** | 0.6732 ± 0.0211 *** | 0.8743 ± 0.0145 *** | 0.7591 ± 0.0137 *** | 0.8620 ± 0.0153 *** | 56 min | 79 |
| AG_Res_UNet | 8.02 millions | 0.0122 ± 0.0006 *** | 0.7631 ± 0.0155 *** | 0.9679 ± 0.0012 *** | 0.6765 ± 0.0207 *** | **0.8758 ± 0.0188** *** | 0.7617 ± 0.0148 *** | 0.8676 ± 0.0164 *** | 1 h 27 min | 59 |
| SE_Res_UNet | 8.81 millions | 0.0122 ± 0.0005 *** | 0.7655 ± 0.0175 *** | 0.9681 ± 0.0011 *** | 0.6826 ± 0.0258 *** | 0.8718 ± 0.0109 *** | 0.7634 ± 0.0161 *** | 0.8656 ± 0.0167 *** | 44 min | 92 |
| TransUNet | 32.6 millions | 0.0123 ± 0.0005 *** | 0.7721 ± 0.0170 *** | 0.9675 ± 0.0011 *** | 0.7214 ± 0.0271 *** | 0.8309 ± 0.0081 *** | 0.7657 ± 0.0160 *** | 0.8607 ± 0.0169 *** | 1 h 36 min | 65 |
| Pix2pix | 30.6 millions | 0.0179 ± 0.0007 *** | 0.6889 ± 0.0170 *** | 0.9534 ± 0.0014 *** | 0.6805 ± 0.0158 *** | 0.6979 ± 0.0263 *** | 0.6767 ± 0.0170 *** | 0.7434 ± 0.0220 *** | 1h 09 min | 90 |
| ASE_Res_UNet | **7.96 millions** | **0.0116 ± 0.0004** | **0.7825 ± 0.0156** | **0.9688 ± 0.0011** | **0.7345 ± 0.0233** | 0.8377 ± 0.0108 | **0.7762 ± 0.0148** | **0.8730 ± 0.0147** | **36 min** | **99** |

**Table 4: ASE_Res_UNet outperforms other advanced architectures in microtubule segmentation on the *MicSim_FluoMT complex* dataset.**
Microtubule segmentation performances of ASE_Res_UNet and other advanced architectures on the *MicSim_FluoMT complex* dataset, evaluated using various metrics (mean ± standard deviation over 120 test images). Bold values indicate the best performance metrics, the lowest model parameter number, the shortest training time, or the best inference rate. Statistical differences between ASE_Res_UNet and other advanced architectures are all significant (Student t-test per metric: ***: $p \leq 0.0001$).

Second, we investigated whether architectures based on fundamentally different design paradigms could outperform ASE_Res_UNet in segmenting microtubules in noisy fluorescence images. We selected an architecture with a Generative Adversial Network (GAN) module alongside a U-Net, i.e. pix2pix model [91] (Fig. S5D). GAN-based methods have previously shown superior performance over standard U-Net or Residual U-Net architectures in tasks such as retinal vessel or corneal nerve segmentation [71, 87-89]. We also included TransUNet, an architecture that integrates transformer layers with a U-Net [81, 92] (Fig. S5E), recently applied to various biomedical image segmentation tasks [81, 92-95]. While Pix2pix showed poor performance, TransUNet produced results closer to ASE_Res_UNet, though at the cost of having approximately four times more parameters (Tables 4 and S5). Specifically, TransUNet had a true-positive count most similar to that of ASE_Res_UNet, albeit with slightly higher false-positive and false-negative counts (Table S5). Visually, Pix2pix produced predictions with more hallucinated structures and shorter microtubules (Fig. 4G), consistent with its lower and higher numbers of true positives and false positives, respectively (Table S5). In contrast, the predictions from TransUNet closely resembled those of ASE_Res_UNet, particularly in the accurate segmentation of microtubule extremities (Fig. 4H). These results suggested that the GAN-based approach was not well suited for segmenting microtubules in noisy images. While the transformer-based TransUNet architecture showed promising performance, it remained inferior to ASE_Res_UNet, required substantially more computational resources and achieved slower training and inference speeds.

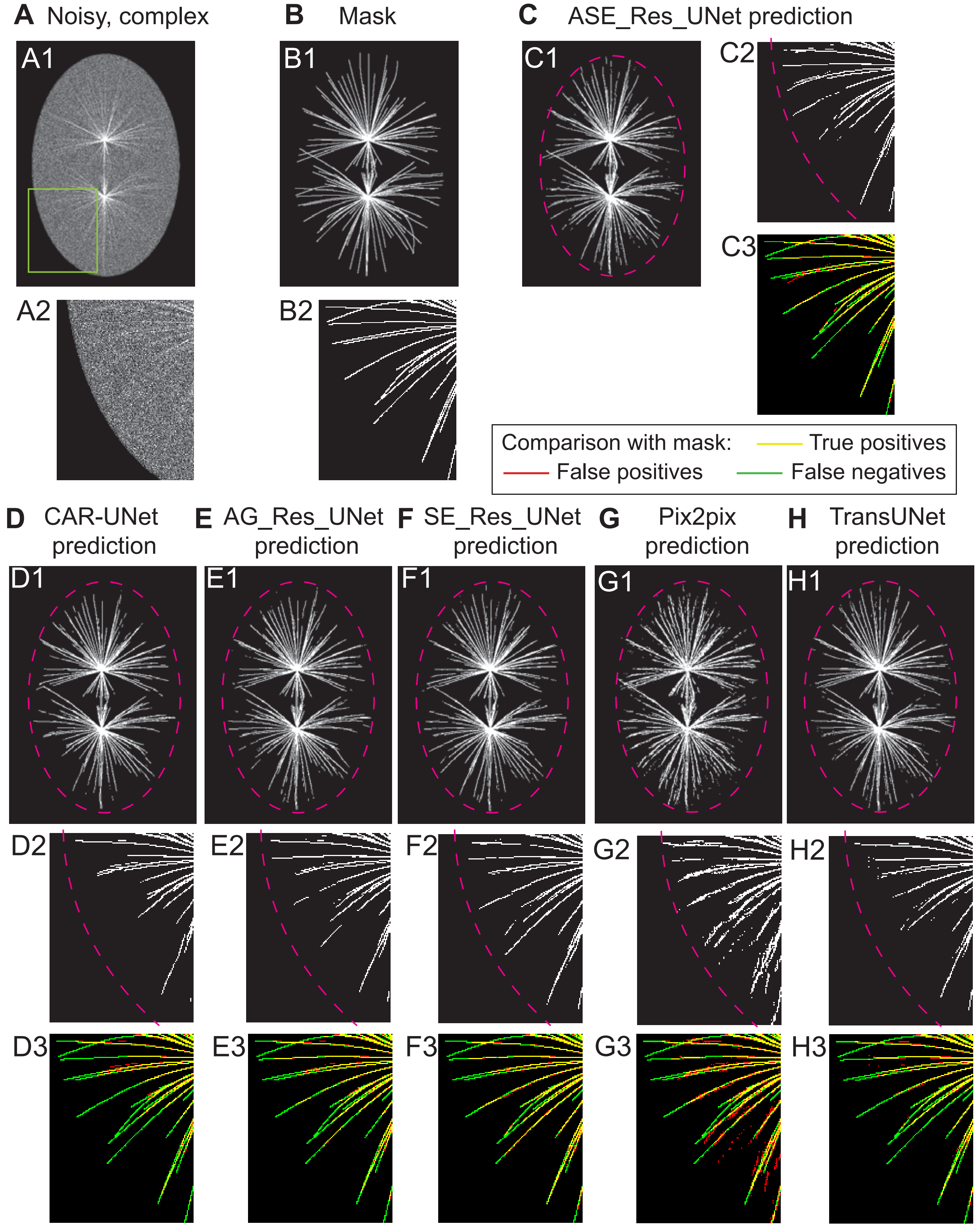

**Fig. 4: ASE_Res_UNet produces the most accurate segmentation predictions compared to state-of-the-art architectures.**

Microtubule segmentation results obtained using ASE_Res_UNet and five other advanced architectures on a sample test image from the *MicSim_FluoMT complex* dataset. (**A**) Input image; (**B**) corresponding ground truth; (**C-H**) predicted segmentations from (C) ASE_Res_UNet, (D) CAR-UNet, (E) AG_Res_UNet, (F) SE_Res_UNet, (G) Pix2pix, and (H) TransUNet models. (A1-H1): whole simulated embryo; (A2-H2): zoomed-in regions of interest (ROI) to better highlight differences between model predictions and ground truth, with cropping window shown in green on panel A1; (C3-H3) composite ROI images showing true positives in yellow, false negatives in green, false positives in red, and true negatives in black. Dashed pink line indicates the embryo contour.

All models occasionally hallucinated structures in the peripheral regions, where microtubule intensity approached background noise. To assess this systematically, we computed performance metrics separately for central vs. peripheral regions (Table S6). ASE_Res_UNet consistently outperformed all others in both regions, with the largest margin in the periphery. TransUNet was the second-best performer, but with significantly reduced precision, indicating more hallucinations. AG_Res_UNet achieved the highest precision overall (fewest false positives), but underperformed in peripheral detection due to higher false negatives.

ASE_Res_UNet provided the best balance between detecting low-intensity microtubule extremities and minimising hallucinations, while also maintaining the lowest computational cost. These results confirmed that ASE_Res_UNet is well-suited for challenging segmentation tasks involving noisy, low-contrast fluorescence microscopy images.

### 3.4 ASE_Res_UNet achieves near-perfect microtubule segmentation on the SynthMT dataset

To further evaluate ASE_Res_UNet, we tested its performance on the SynthMT dataset [31], an independent synthetic microtubule dataset that differs from *MicSim_FluoMT* in two key aspects (cf. section 2.1): it features individual *in vitro* microtubules (rather than *in vivo* cytoskeletal networks) and exhibits a higher SNR. Additionally, this dataset provides a large training set (5280 images). Given these characteristics, we foresaw that microtubule segmentation would be comparatively easier, yielding strong performance. Visual inspection confirmed this hypothesis: ASE_Res_UNet's predictions closely aligned with ground-truth masks (Fig. 5). Quantitative metrics further validated its accuracy: Dice = 0.9807 $\pm$ 0.0171, IoU = 0.9981 $\pm$ 0.0020, Sensitivity = 0.9854 $\pm$ 0.0099, Precision = 0.9764 $\pm$ 0.0282, MCC = 0.9803 $\pm$ 0.0172, and PR AUC = 0.9985 $\pm$ 0.0024 (mean $\pm$ SD over 660 test images). ASE_Res_UNet outperformed foundation models from the SynthMT study, including the recently introduced SAM3Text (average precision of 0.95) [31]. These results demonstrated that ASE_Res_UNet, originally developed for noisy microtubule segmentation, also excels in high-SNR conditions.

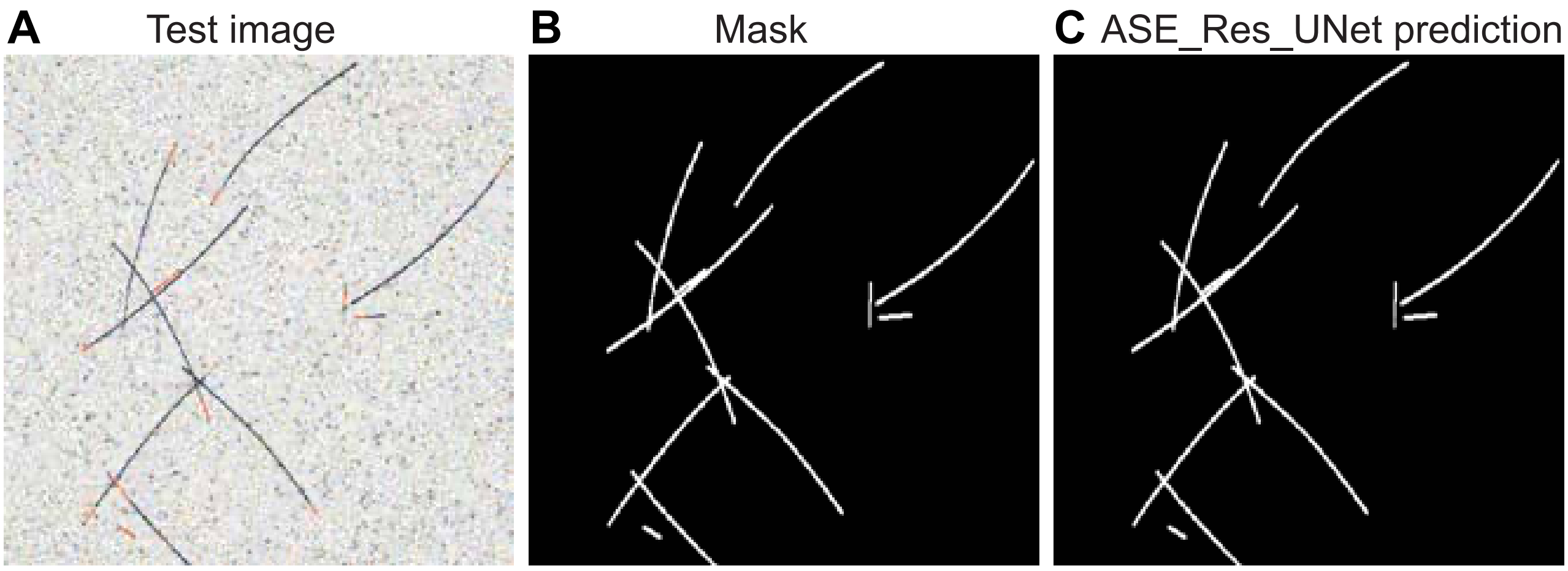

**Fig. 5: ASE_Res_UNet predictions closely match ground-truth masks in the SynthMT dataset**
(**A**) Exemplar synthetic test image, (**B**) Ground-truth mask, and (**C**) ASE_Res_UNet prediction (model trained on SynthMT).

### 3.5 ASE_Res_UNet enables to segment microtubules in real microscopy images.

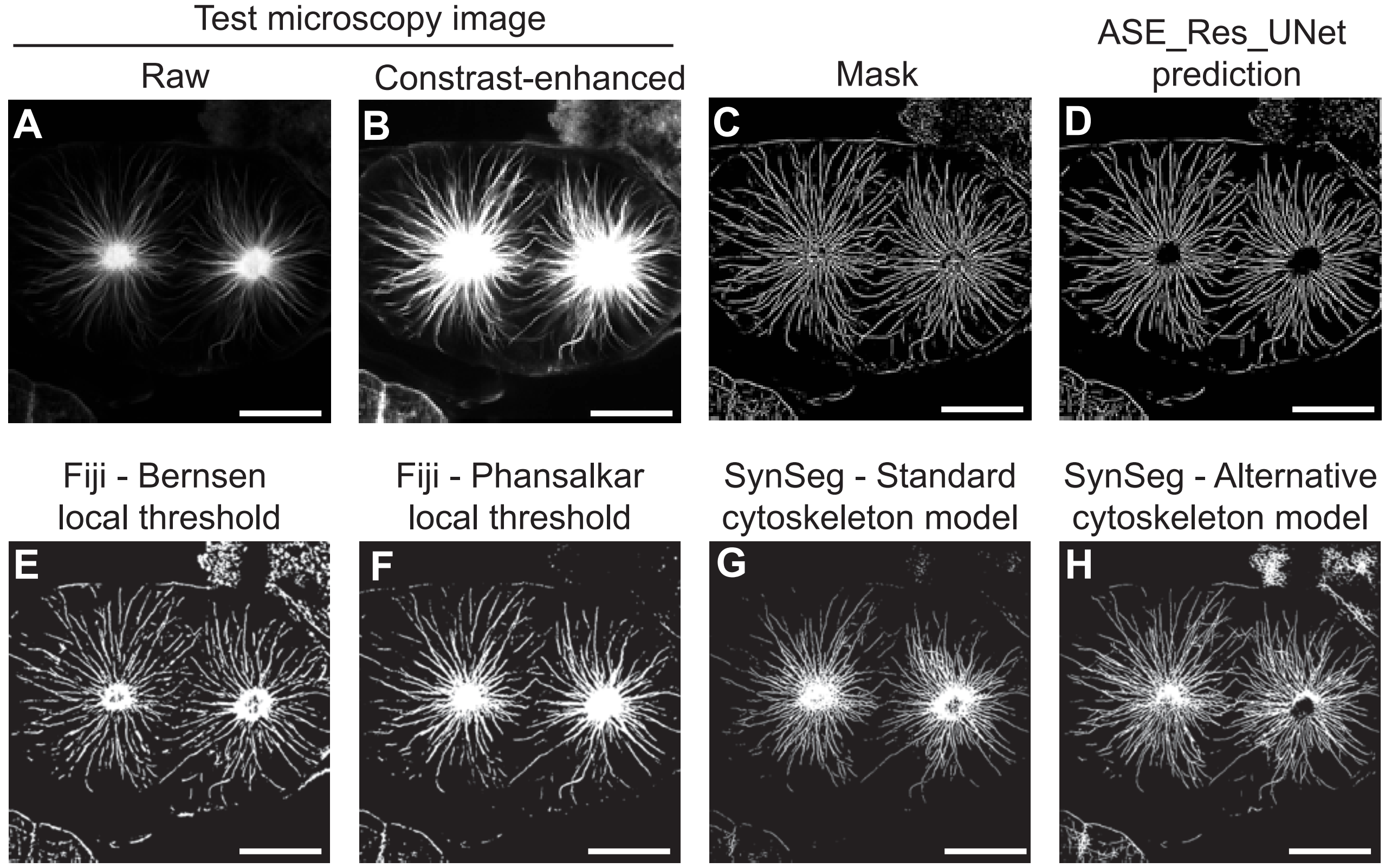

**Fig. 6: ASE_Res_UNet successfully segments microtubules in fixed embryos, including dense centrosomal regions and faint peripheral areas.**
Microtubule segmentation in a fixed *C. elegans* zygote stained with anti-tubulin antibody: (**A, B**) Test microscopy image: (A) raw, and (B) contrast-enhanced (to see low-contrast filaments and filament extremities), (**C**) Ground-truth mask, (**D**) ASE_Res_UNet prediction (model trained on the *MicReal_FluoMT* dataset), (**E, F**) Fiji-based segmentations using local thresholding: (E) Bernsen or (F) Phansalkar, and (**G, H**) SynSeg predictions from pre-trained cytoskeleton models with smart auto-resize: (G) standard model, and (H) alternative model. Scale bar indicates 10 $\mu$m.

We next questioned whether ASE_Res_UNet would maintain its high performance on real microscopy images, which often pose additional challenges not fully captured in synthetic datasets. Real images may exhibit biological and experimental variability, imaging artifacts, or background structures that complicate learning. Moreover, the limited size of annotated real-world datasets—due to the difficulty and time cost of manual annotation—can further hinder training. To address this, we evaluated ASE_Res_UNet on the *MicReal_FluoMT* dataset, which contains 49 fluorescence microscopy images of *C. elegans* fixed embryos stained with anti-tubulin antibodies (cf. Section 2.2). This dataset represents a realistic biological context with naturally occurring imaging noise, variability in filament orientation and density, and heterogeneous fluorescence intensities (e.g., bright at centrosomes and faint at the cell periphery). We trained ASE_Res_UNet on 19 images from the *MicReal_FluoMT* dataset, reserving 10 images for validation and 10 for testing. Despite the limited training data, ASE_Res_UNet accurately segmented astral microtubules (Fig. 6). Notably, it successfully detected microtubule extremities that were barely visible in the raw images (Fig. 6A, C), and captured filaments with weak or uneven staining. Interestingly, the model produced segmentations that were often smoother and more continuous than the corresponding manual annotations (Fig. 6C vs. 6B), suggesting that the model may surpass manual annotations in terms of consistency and completeness. Quantitative metrics supported these observations: IoU = 0.9010 $\pm$ 0.0284, accuracy = 0.9477 $\pm$ 0.0158, specificity = 0.9823 $\pm$ 0.0070, precision = 0.7795 $\pm$ 0.0604, and PR AUC = 0.7779 $\pm$ 0.0502 (mean $\pm$ standard deviation over 10 test images). We compared ASE_Res_UNet predictions with segmentations generated using standard image-processing tools in Fiji, a widely adopted software in biological research [96]. Among local thresholding methods, the Bernsen algorithm (suitable for edge detection [97]) and the Phansalkar algorithm (optimised for low contrast images [98]) achieved the highest performance (Fig. 6E-F). However, these methods produced shorter microtubules and omitted some microtubules compared to ASE_Res_UNet. Overall, ASE_Res_UNet demonstrated robust segmentation performance in fixed *C. elegans* embryos, effectively capturing microtubules in both dense and low-signal regions, despite the limited size of the training dataset. These results highlighted its potential as a valuable tool for biologists, offering high-quality and efficient segmentation of cytoskeletal filaments in fluorescence microscopy images.

To further evaluate our method's performance, we compared ASE_Res_UNet with SynSeg, a state-of-the-art deep-learning framework for subcellular structure segmentation—including microtubules—that employs a synthetic data-driven approach [57]. SynSeg offers two pre-trained models for cytoskeleton segmentation (termed standard and alternative), which we benchmarked on the *MicReal_FluoMT* dataset. Visual inspection revealed that both SynSeg models produced lower-quality segmentations than ASE_Res_UNet, with the standard model performing particularly poorly (Fig. 6 G, H). SynSeg's outputs exhibited shorter microtubules, discontinuous filaments, and missed detections. Quantitative metrics further confirmed ASE_Res_UNet's superiority. The alternative SynSeg model achieved an IoU of 0.8641 $\pm$ 0.0350, accuracy of 0.9268 $\pm$ 0.0201, specificity of 0.9728 $\pm$ 0.0096, precision of 0.6251 $\pm$ 0.0430, and PR AUC of 0.5705 $\pm$ 0.0436 (mean $\pm$ standard deviation over 10 test images). Collectively, these findings established ASE_Res_UNet as a highly competitive—if not superior—approach for microtubule segmentation compared to this cutting-edge framework.

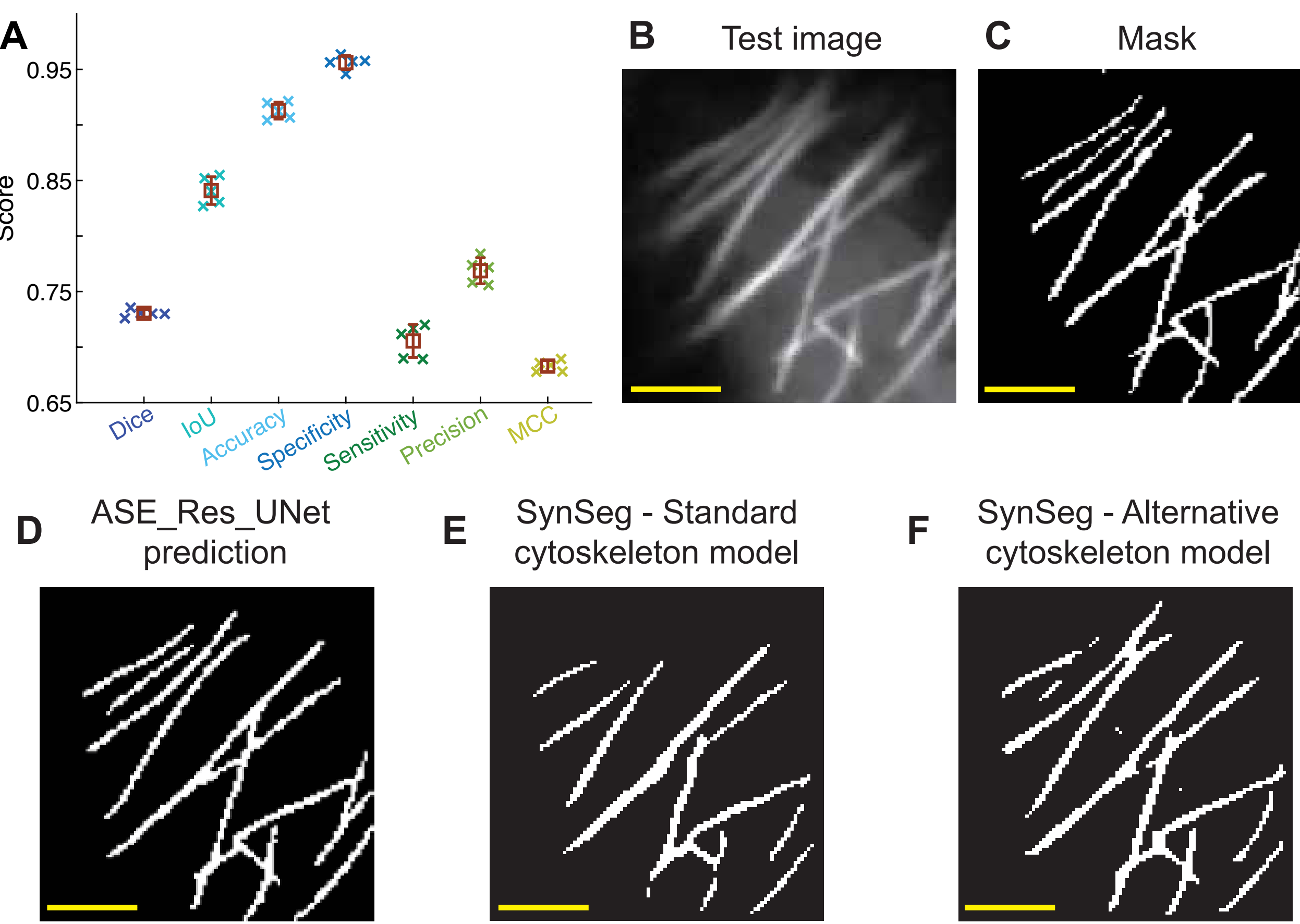

**Fig. 7: ASE_Res_UNet accurately segments microtubules in *Nicotiana tabacum* BY-2 cells.**
(**A**) Performance metrics for ASE_Res_UNet across 5-fold cross-validation (CV) on 90 re-annotated images from the Higaki dataset. Crosses represent individual fold score, while brown squares and error bars indicate the means $\pm$ standard deviation across fold. (**B**) Test image (unseen during CV), (**C**) corresponding mask (re-annotated by Guo *et al*.), (**D**) ASE_Res_UNet prediction (trained on a given fold), (**E, F**) predictions from SynSeg pre-trained cytoskeleton models with smart auto-resize: (E) standard model, and (F) alternative model. Yellow scale bar indicates 5 μm.

We next evaluated ASE_Res_UNet on a previously published confocal microtubule image dataset (Higaki dataset) that was recently partially reannotated by Guo *et al.* [57] (cf. section 2.2). Using 90 images of *Nicotiana tabacum* BY-2 cells, we performed five-fold cross-validation and achieved strong performances (Fig. 7A-D). ASE_Res_UNet outperformed SynSeg predictions from pretrained cytoskeleton models (Fig. 7E, F), which yielded an averaged precision of ~0.65 and an averaged Dice of ~0.70 (see Fig. 6 in [57]). These results demonstrated ASE_Res_UNet's competitiveness in segmenting microtubules in real images with diverse filament networks and varied microscopy modalities.

To further assess ASE_Res_UNet's cross-domain robustness under a significant imaging-modality shift, we applied the model trained on *MicReal_FluoMT* to live *C. elegans* mitotic embryos expressing GFP::TBB-2, without fine-tuning, transfer learning, or preprocessing. Since astral microtubules are very dynamic in the *C. elegans* zygote, it calls for high-speed imaging, leading to low SNR and challenging segmentation conditions (Fig. 8A). ASE_Res_UNet successfully segmented most microtubules along their entire length, demonstrating strong generalisation (Fig. 8B). However, faint microtubules were detected with lower accuracy, suggesting that semi-supervised learning with partial annotations could improve performance in such cases. ASE_Res_UNet outperformed both the Bernsen and Phansalkar local thresholding algorithms (Fig. 8C, D), and SynSeg predictions from standard and alternative cytoskeleton models (Fig. 8E, F). Notably ASE_Res_UNet improved microtubule continuity compared to SynSeg's alternative cytoskeleton model. These results suggested that ASE_Res_UNet holds promise for applications across diverse biological samples, including dynamic imaging conditions.

Overall, ASE_Res_UNet performed robustly on real microscopy images, including datasets with different imaging modalities and cell types. It offered a novel tool for biological image analysis, enabling accurate and efficient microtubule segmentation across a range of experimental conditions.

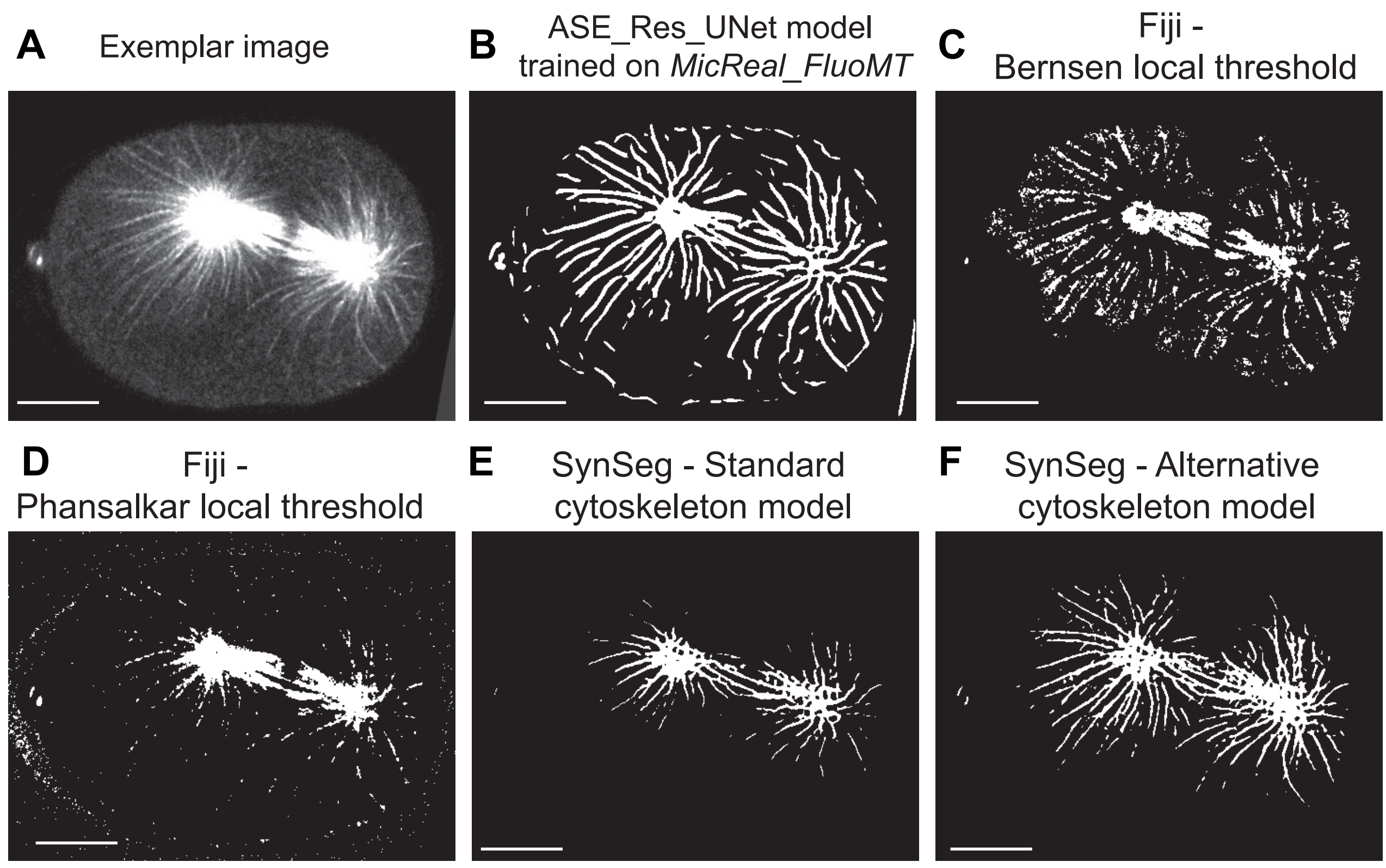

**Fig. 8: ASE_Res_UNet trained on the *MicReal_FluoMT* dataset generalizes to live *C. elegans* embryos.**

Microtubule segmentation in a *C. elegans* zygote expressing GFP::β-tubulin: (**A**) exemplar fluorescence image, (**B**) ASE_Res_UNet prediction (model trained on the *MicReal_FluoMT* dataset without fine-tuning or transfer learning), (**C, D**) segmentations via Fiji's local thresholding methods: (C) Bernsen and (D) Phansalkar, and (**E, F**) predictions from SynSeg pre-trained cytoskeleton models with smart auto-resize: (E) standard model, and (F) alternative model. Scale bar indicates 10 μm.

### 3.6 Strong transferability of ASE_Res_UNet for segmenting vessels and nerves in biomedical images

Finally, we assessed ASE_Res_UNet's ability to segment curvilinear structures beyond microtubules. First, we evaluated the model on the *DRIVE* dataset, which contains retinal fundus images with annotated blood vessels that differ significantly in appearance, scale and imaging modality (cf. section 2.2). ASE_Res_UNet trained on *DRIVE* was able to accurately capture complex vessel patterns, including thin and low-contrast vessels (Fig. 9A-C). Quantitative evaluation further confirmed the model's strong performance (Table 5). To assess competitiveness, we compared ASE_Res_UNet with CAR-UNet, a model specifically designed for retinal vessel segmentation. Despite using approximately half the number of trainable parameters, ASE_Res_UNet achieved slightly better performance (Table 5), and successfully segmented some faint vessels missed by CAR-UNet (Fig. 9D vs. 9C). We extended this comparison to AG_Res_UNet and SE_Res_UNet, which differed from ASE_Res_UNet only in their attention mechanisms. ASE_Res_UNet outperformed both models in terms of precision (Table 5), which was consistent with qualitative observations (Fig. 9E, F). Last, we compared ASE_Res_UNet to TransUNet, which had previously shown competitive performance on the microtubule segmentation task. ASE_Res_UNet again slightly outperformed TransUNet on the DRIVE dataset and provided a better balance between false positives and false negatives (Table 5, Fig. 9G). Importantly, ASE_Res_UNet outperformed or was competitive with other methods while requiring significantly fewer computational resources, achieving the fastest inference time and the lowest number of model parameters. Overall, ASE_Res_UNet demonstrated versatility in segmenting curvilinear structures beyond microtubules, achieving state-of-the-art performance on a segmentation task for which it was not originally designed.

| **Model** | **Parameter number** ↓ | **Loss** ↓ | **Dice** ↑ | **IoU** ↑ | **Sensitivity** ↑ | **Precision** ↑ | **MCC** ↑ | **PR AUC** ↑ | **Training time** ↓ | **Inference rate** ↑ (image/s) |
|---|---|---|---|---|---|---|---|---|---|---|
| CAR-UNet | 15.83 millions | 0.1563 ± 0.0210 | 0.8159 ± 0.0165 | 0.9414 ± 0.0059 | 0.8036 ± 0.0427 | **0.8312 ± 0.0314** | 0.8003 ± 0.0171 | 0.9032 ± 0.0122 | **5 min** | 110 |
| AG_Res_UNet | 8.02 millions | 0.1601 ± 0.0216 | 0.8171 ± 0.0158 | 0.9401 ± 0.0053 | 0.8289 ± 0.0472 | 0.8084 ± 0.0295 | 0.8011± 0.0162 | 0.9024 ± 0.0147 | **5 min** | 97 |
| SE_Res_UNet | 8.81 millions | 0.1636 ± 0.0218 | 0.8174 ± 0.0191 | 0.9414 ± 0.0065 | 0 .8117 ± 0.0417 | 0.8254 ± 0.0286 | 0.8016 ± 0.0204 | 0.9019 ± 0.0171 | 6 min | 102 |
| TransUNet | 32.3 millions | 0.1557 ± 0.0227 | 0.8150 ± 0.0129 | 0.9389 ± 0.0049 | **0.8347 ± 0.0378** | 0.7983 ± 0.0291 | 0.7987 ± 0.0138 | 0.9004 ± 0.0152 | 7 min | 85 |
| ASE_Res_UNet | **8.00 millions** | **0.1481 ± 0.0221** | **0.8236 ± 0.0167** | **0.9433 ± 0.0055** | 0.8185 ± 0.0432 | **0.8311 ± 0.0294** | **0.8084 ± 0.0178** | **0.9114 ± 0.0150** | **5 min** | **126** |

**Table 5: ASE_Res_UNet achieves state-of-the-art performance on segmenting retinal blood vessels.** Retinal vessel segmentation performances of ASE_Res_UNet and other advanced architectures on the *DRIVE* dataset, evaluated using various metrics (mean ± standard deviation over 20 test images). Bold values indicate the best performance metrics, the lowest model parameter number, the shortest training time, or the best inference rate. Statistical differences between ASE_Res_UNet and other architectures are non-significant.

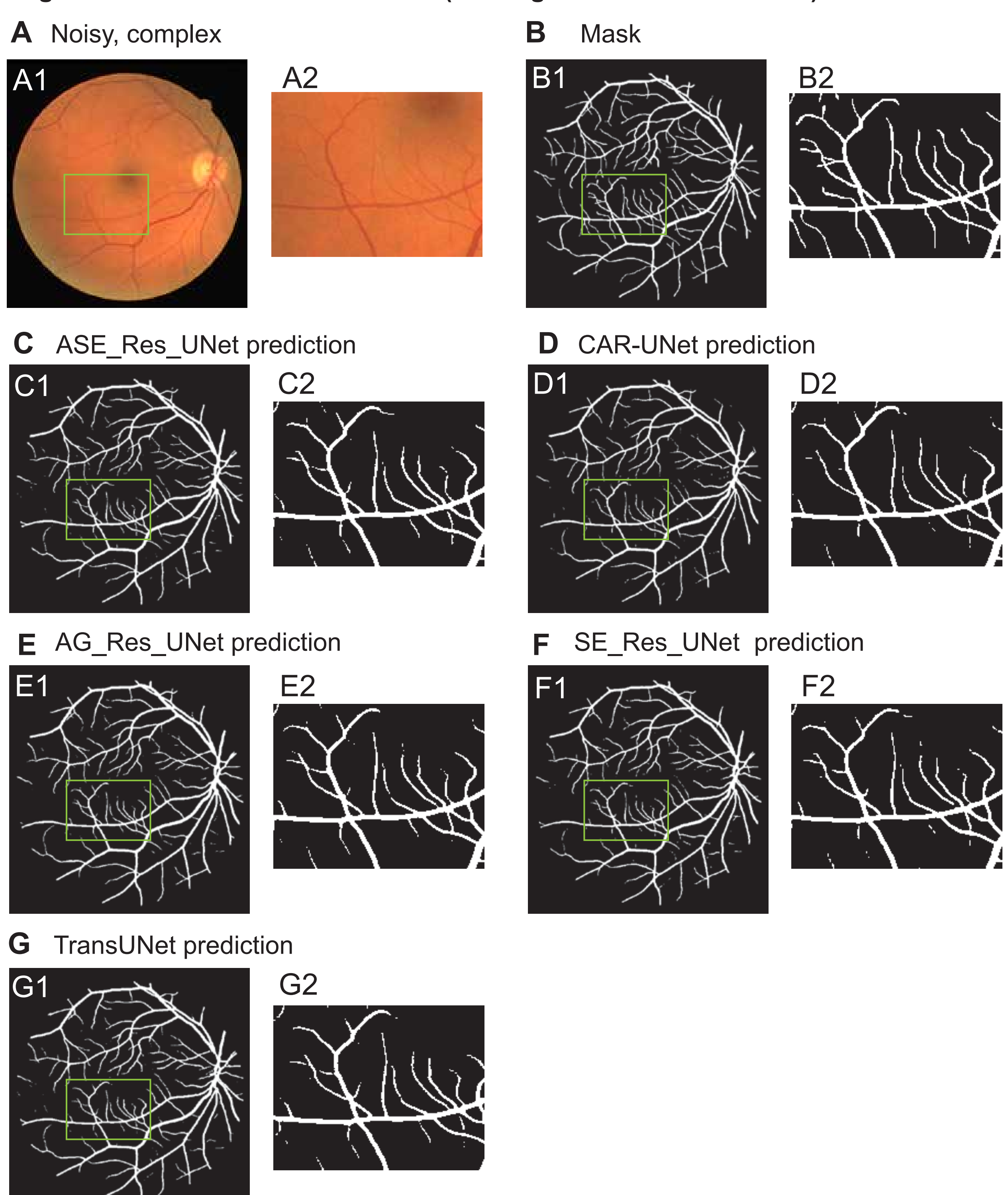

**Fig. 9: ASE_Res_UNet successfully segments retinal blood vessels, including thin and low-contrast vessels, while limiting false positives.**

Retinal vessel segmentation results obtained using ASE_Res_UNet and four advanced architectures on a sample test image from the *DRIVE* dataset. (**A**) Input image; (**B**) corresponding ground truth; (**C-G**) predicted segmentations from (C) ASE_Res_UNet, (D) CAR-UNet, (E) AG_Res_UNet, (F) SE_Res_UNet, and (G) TransUNet models. (A1-G1): whole images; (A2-G2): zoomed-in regions of interest (ROI) to better highlight differences between model predictions and ground truth, with cropping window shown in green on panels A1 and B1.

To further assess the data-wise robustness of our approach, we evaluated ASE_Res_UNet on the *CORN-1* dataset, which contains fluorescence microscopy images of corneal nerves (cf. section 2.2). ASE_Res_UNet model that was 5-fold cross-trained on *CORN-1* successfully segmented the nerve structures, including those with weak fluorescence signals (Fig. 10C vs. 10B), and it performed better than U-Net in this task (Fig. 10D). Quantitative metrics aligned with these findings: ASE_Res_UNet outperformed U-Net across all metrics (Table 6). Its success on three distinct curvilinear structure segmentation tasks—microtubules, blood vessels, and corneal nerves—underscored the robustness and adaptability of ASE_Res_UNet's architectural design. In particular, the combination of residual encoding and noise-adaptive ASE attention modules enables effective feature extraction and denoising across diverse biomedical imaging contexts.

| Model | Loss ↓ | Dice ↑ | IoU ↑ | Sensitivity ↑ | Precision ↑ | MCC ↑ | PR AUC ↑ |
|---|---|---|---|---|---|---|---|
| U-Net | 0.0416 ± 0.0017<br>◊ | 0.6993 ± 0.0123<br>* | 0.9837 ± 0.0004<br>◊ | 0.6647 ± 0.0267<br>◊ | 0.7523 ± 0.0135 | 0.7023 ± 0.0112<br>* | 0.7895 ± 0.0107<br>◊ |
| ASE_Res_UNet | **0.0370 ± 0.0026** | **0.7298 ± 0.0140** | **0.9851 ± 0.0008** | **0.7015 ± 0.0135** | **0.7726 ± 0.0200** | **0.7321 ± 0.0138** | **0.8214 ± 0.0176** |

**Table 6: ASE_Res_UNet outperforms U-Net for segmenting corneal nerves.**
Retinal nerve segmentation performances of ASE_Res_UNet and U-Net on the CORN-1 dataset, evaluated using various metrics (mean ± standard deviation over the 5-fold cross-training). Bold values indicate the best performances. Statistical differences between ASE_Res_UNet and U-Net are indicated only when significant (◊: $0.01 < p \leq 0.05$; *: $0.001 < p \leq 0.01$).

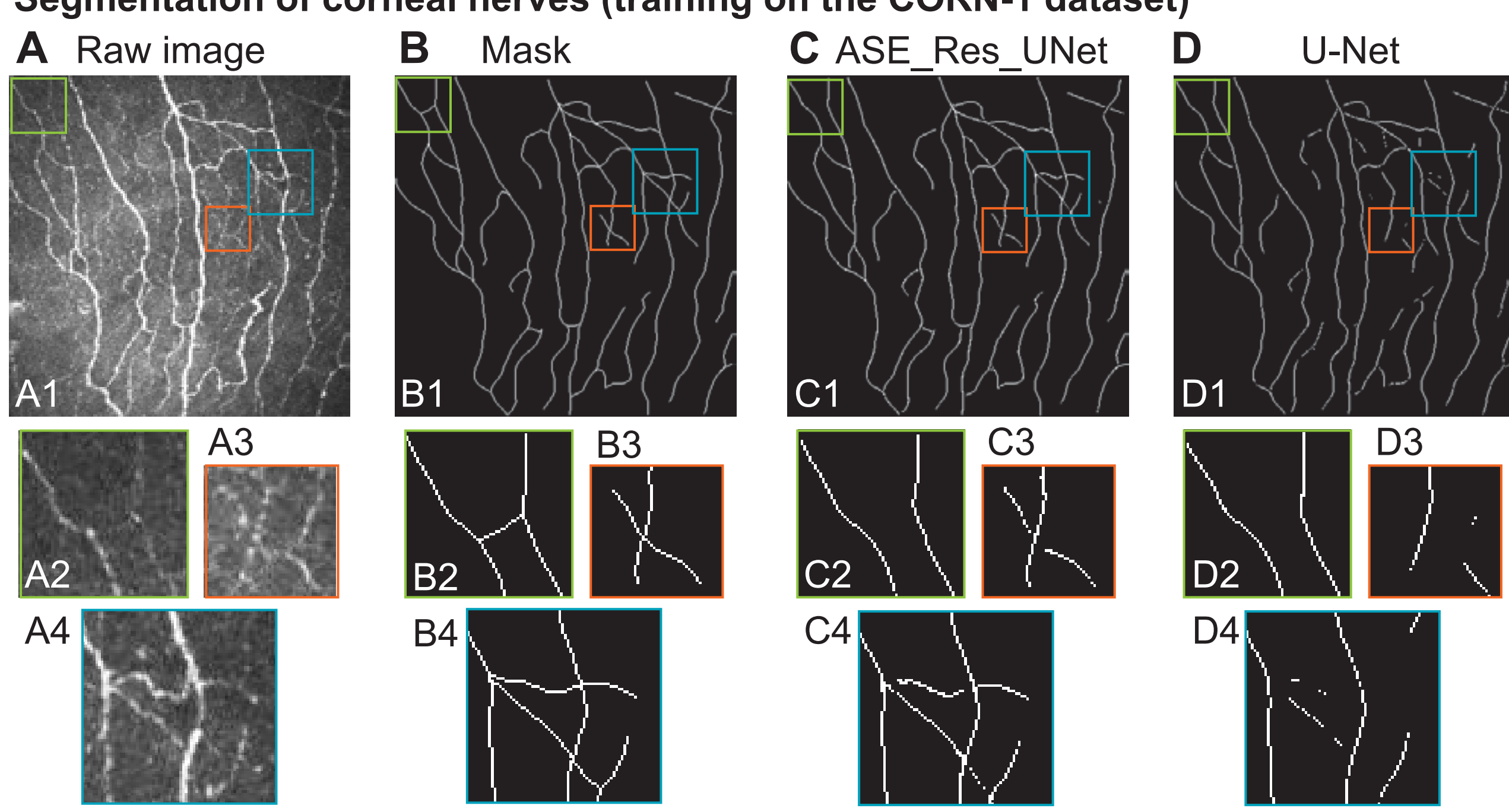

**Fig. 10: ASE_Res_UNet successfully segments corneal nerves in noisy fluorescence microscopy images.**

Retinal nerve segmentation results obtained using ASE_Res_UNet and U-Net trained on the *CORN-1* dataset. (**A**) Exemplar test image; (**B**) corresponding ground truth; (**C-D**) predicted segmentations from (C) ASE_Res_UNet, and (D) U-Net models. (A1-D1): whole image; (A2-D2): first zoomed-in region of interest (ROI) indicated in orange; (A3-D3) second zoom-in ROI indicated in green; and (A4-D4) third zoom-in ROI indicated in blue.

## 4 Discussion

This work positions ASE_Res_UNet as a novel computationally efficient tool for the segmentation of microtubules in fluorescence microscopy, even under challenging conditions such as background fluorescence noise, inhomogeneous backgrounds, low contrast, and uneven fluorescence intensity. On the *MicSim_FluoMT easy* dataset, it successfully handles background fluorescence often present in microscopy (Table 2; Fig. 2B and S3F). On the *MicSim_FluoMT complex* dataset, it segments filaments with low SNR, as encountered in live-cell imaging (Table 3; Fig. 2C, 3F, 4C). On the *MicReal_FluoMT* dataset and Higaki reannotated dataset, it effectively segments filaments with varying contrast and fluorescence (Fig. 6 & 7). ASE_Res_UNet also demonstrates strong versatility by successfully segmenting distinct curvilinear structures, such as retinal blood vessels (*DRIVE* dataset) and retinal nerves (*CORN-1* dataset), despite differences in imaging modalities and structural characteristics (Tables 5, 6; Fig. 9C, 10C). This highlights model's applicability beyond microtubule segmentation.

A key innovation is the Adaptive Squeeze-and-Excitation (ASE) mechanism, which explicitly incorporates image noise into the attention process. This improves performance, especially in noisy conditions. Indeed, ASE_Res_UNet outperforms SE_Res_UNet on the *MicSim_FluoMT complex* dataset (Table 4; Fig. 4C, F) and produces cleaner segmentations on the *DRIVE* dataset (Table 5; Fig. 9C, F). These gains—particularly in sensitivity and false-positive reduction—can be attributed to its noise-adaptive decoding. The ablation study further reveals the complementary strengths of the model's architectural components (Tables 3 and S3; Fig. 3). Residual blocks enhance gradient flow and support the detection of densely packed filaments. Meanwhile, the ASE attention mechanism proves particularly beneficial in peripheral regions, where microtubule extremities appear with low intensity and are more prone to misclassification (Table S4). However, we observe a trade-off between sensitivity and precision: the increase in sensitivity—driven by the noise-adaptive mechanism—is accompanied by a decrease in precision. This reflects an inherent limitation of the current implementation, where the model's ability to capture faint or noisy structures sometimes results in false positives. While mitigating this trade-off warrants further investigation, the primary focus of this work was to demonstrate the effectiveness of the noise-adaptive mechanism in improving sensitivity for challenging imaging conditions.

By segmenting microtubules in noisy images, we addressed two main challenges; class imbalance and annotation quality. Microtubules often occupy a small fraction of an image, leading to a dominance of background pixels. While ASE_Res_UNet trained on the *MicSim_FluoMT complex* dataset achieved high scores on conventional metrics (accuracy = 0.9842 ± 0.0006, specificity = 0.9942 ± 0.0005, ROC AUC = 0.9911 ± 0.0011), we observed that these metrics fail to capture segmentation quality at filament extremities. Therefore, we recommend using Dice, sensitivity, precision, MCC, and PR AUC, which better align with visual inspection. We also found that weighted cross-entropy (WCE), which gives more weight to underrepresented classes, performs better than other loss functions (Table 1). Although effective, WCE remains underused in the literature [53, 69, 82, 84]. The second challenge was the limited availability of high-quality annotations for microtubules in noisy fluorescence images. We addressed this by generating synthetic training data, which offers two advantages: (i) controlled variation of imaging parameters, and (ii) perfect ground truth masks for reliable metric evaluation. Comparing performance on the *MicSim_FluoMT* easy and complex synthetic datasets reveals how the model adapts to different imaging challenges (Tables 2–3; Fig. 2). The performance of models trained on real datasets can be affected by inaccuracies or noise in the annotations, as observed in the *MicReal_FluoMT* and *CORN-1* datasets. In the *CORN-1* dataset, annotations of nerve structures may vary between experts. For example, in Fig. 10A2, a barely visible nerve is annotated in the center of the image. Another expert might conclude that this structure does not exist, implying that its omission by ASE_Res_UNet in Fig. 10C2 could be correct.

Consequently, counting this as a false negative would underestimate the model's sensitivity. In the *MicReal_FluoMT* dataset, imperfect annotations arise from the semi-automated pipeline used to generate them. This pipeline occasionally mislabels background noise as microtubules (Fig. 6C). ASE_Res_UNet effectively filters out such noise by excluding these pixels from its predictions (Fig. 6D), but this results in false negatives, thereby underestimating sensitivity. Conversely, the pipeline sometimes fails to capture the continuity of microtubules in its annotations. When ASE_Res_UNet correctly identifies these structures, they are counted as false positives, which underestimates precision. These observations underscore the challenges associated with real-world annotations and highlight the potential advantages of synthetic datasets, which provide perfectly aligned ground-truth masks without annotation ambiguities. Although synthetic data may raise concerns about model robustness, our results demonstrate that ASE_Res_UNet, initially developed using synthetic datasets, performs effectively across real-world curvilinear tasks when trained on corresponding real datasets (Fig. 6, 7, 9 & 10). However, the direct application of a model trained exclusively on synthetic data to real images remains challenging, particularly in noisy scenarios. This is because the model learns a specific noise pattern from synthetic data, whereas real-world images exhibit noise characteristics that vary with the microscope, imaging conditions, fluorophore, and sample treatment, all of which influence noise distribution and intensity. Interestingly, ASE_Res_UNet, when trained on fixed-embryo images (*MicReal_FluoMT* dataset), produces promising qualitative results on live-embryo images without retraining (Fig. 8). Future work will explore transfer learning, hybrid training strategy, and domain-specific adaptation to further improve generalisation performance.

Compared to previous tools for curvilinear structure segmentation—such as SIFNE [28] or SOAX [30], which often require multiple processing steps and manual parameter tuning—ASE_Res_UNet is an end-to-end deep learning solution requiring no handcrafted features or denoising. Unlike many models, it is not limited to a specific structure or imaging modality and remains performant even in noisy conditions. For instance, in contrast to GAN-Based denoising approaches [99, 100], ASE_Res_UNet directly integrates noise adaptation into the attention mechanism, avoiding the need for separate preprocessing steps. Additionally, its lightweight architecture enables deployment in resource-constrained settings and allows training with relatively few annotated images, as shown on the *MicReal_FluoMT* and *DRIVE* datasets. Importantly, ASE_Res_UNet outperforms various models: models with alternative attention mechanisms (CAR-UNet, AG_Res_UNet, SE_Res_UNet), models with distinct architectures (Pix2pix, TransUNet), and a recent microtubule-segmentation framework (SynSeg). Notably, ASE_Res_UNet outperforms CAR-UNet on the *DRIVE* dataset, despite having half the parameters and not being specifically designed for retinal vessel segmentation (Table 5; Figures 9C–D). In addition, ASE_Res_UNet achieves competitive training and inference speeds compared to larger models, as demonstrated on the *MicSim_FluoMT* complex and *DRIVE* datasets (Tables 4 and 5).

## 5 Conclusion

ASE_Res_UNet is a compact, robust and versatile framework for curvilinear structure segmentation across diverse imaging contexts. Its contributions span both basic and translational research. In basic research, ASE_Res_UNet can advance cytoskeletal investigations by enabling reliable segmentation of microtubules, actin, and intermediate filaments, particularly in live-cell microscopy, where noise has previously limited quantitative analysis. This could open new avenues in understanding various cell processes and developing therapies targeting cytoskeletal dysfunction in diseases like cancer or neurodegenerative disorders [6-10]. For instance, ASE_Res_UNet could accelerate high-throughput drug screening by automating cytoskeletal phenotype quantification in noisy live-cell images. In clinical applications, ASE_Res_UNet holds promise for disease diagnosis. For instance, retinal vessel tortuosity is a biomarker for conditions like hypertension, diabetic retinopathy, and cardiovascular disease [101-104], while abnormal cerebral vasculature patterns may assist in early detection of brain tumours [105]. By enabling accurate segmentation in challenging conditions, ASE_Res_UNet bridges foundational research and clinical applications, offering tools for biomedical discovery and diagnostic innovation. Finally, our two novel microtubule segmentation datasets—*MicSim_FluoMT* (synthetic) and *MicReal_FluoMT* (real)—are publicly available to enable community-wide benchmarking. These datasets complement recent efforts—the SynthMT dataset [31] and the Higaki 2024 reannotated dataset [57]—by expanding the range of imaging conditions. These resources provide a comprehensive benchmarking framework to advance noise-robust segmentation methods in microscopy.

## Supplementary material

Supplementary material will be available online.

## Acknowledgments

We thank Drs Guangshuo Ou and Zhengyang Guo for generously sharing the Higaki 2024 reannotated dataset [57], which was instrumental in evaluating our model's performance. We thank S. Dutertre and X. Pinson of the Microscopy Rennes Imaging Center (MRic, BIOSIT, Biogenouest) for their assistance. MRic is part of the national infrastructure France-BioImaging supported by the French National Research Agency (ANR-10-INBS-04). We thank Drs. Jacques Pécréaux, Christophe Heligon, Sidi M. Sid'El Moctar, Thierry Pécot, Charles Kervrann, and Florent Autrusseau for helpful discussions about the project. We thank Pécréaux lab and TIAD lab for their support.

## Author contributions

Achraf Ait Laydi: Writing – review & editing, Visualization, Validation, Software, Methodology, Investigation, Formal analysis, Data curation. Louis Cueff: Validation, Software, Methodology, Investigation, Data curation**.** Mewen Crespo: Software, Methodology, Data curation**.** Yousef El Mourabit: Writing – review & editing, Writing – original draft, Validation, Supervision, Software, Project administration, Methodology, Investigation, Funding acquisitions, Data curation, Conceptualization. Hélène Bouvrais: Writing – review & editing, Writing – original draft, Visualization, Validation, Supervision, Software, Resources, Project administration, Methodology, Investigation, Funding acquisitions, Formal analysis, Data curation, Conceptualization.

## Funding

This work was supported by PHC Toubkal 2024 (n° 49945RE). H.B. was supported by the Agence Nationale de la Recherche (JCJC project MICENN, ANR-22-CE45-001601) and the University of Rennes (Défis scientifiques, 2020 ; Soutien aux collaborations internationales, 2024). L.C. was supported by a PhD fellowship from "La Ligue Nationale Contre le Cancer" (2021-2024). The servers used for the computations were funded by the Brittany region (AAP PME 2018-2019 - Roboscope) and by the Agence Nationale de la Recherche" (PRCE project SAMIC, ANR-19-CE45-0011). This project was provided with computing AI and storage resources by GENCI at IDRIS thanks to the grants 2024-AD010315542 and 2025-AD010315542R1 on the supercomputer Jean Zay's V100 partition.

## Data availability.

Original microtubule datasets (synthetic and real noisy images) are available on Zenodo (DOIs: 10.5281/zenodo.14696279 and 10.5281/zenodo.15852660). ASE_Res_UNet model is also shared on Zenodo (DOI: 10.5281/zenodo.17456868).

## Declarations

### Ethics approval and consent to participate

Not applicable.

### Consent for publication

Not applicable.

### Competing interests

Authors declare no conflict of interest.

## Supplementary Material

Article untitled “A novel attention mechanism for noise-adaptive and robust segmentation of microtubules in microscopy mages” by Ait Laydi *et al.* (2026).

# 1- Supplemental text

A- <u>Generation of two synthetic datasets of fluorescence microtubules, called *MicSim_FluoMT*</u>

To address the challenges of segmenting microtubules in noisy environments, we needed a dataset comprising hundreds of noisy images to train the neural network. However, generating such a real dataset is difficult, since collecting and annotating large numbers of images is costly and time-consuming, and high noise levels make accurate annotations unfeasible or imprecise. As a solution, we created two synthetic microscopy datasets of fluorescently labelled microtubules, along with corresponding binary masks (ground truth) using a two-step pipeline

*A1. Image dataset generation and annotation procedure*

We developed a two-step pipeline to generate synthetic images along with their corresponding ground-truths. The first step involved simulating the microtubule mitotic network in the *Caenorhabditis elegans* zygote, a well-established model for studying cell division [1]. For this, we used *Cytosim*, a widely used cytoskeletal simulation tool [2]. The simulations were parameterised using microtubule properties measured either *in vivo* or *in vitro*. We controlled key variables such as microtubule length, curvature, and density to reflect a range of biological scenarios (see table below. This approach allowed us to reproduce an astral microtubule network closely resembling that of the *C. elegans* embryo (Fig. S1A).

**Table:** Objects and their parameters of *Cytosim* simulations

| Object type | Characteristic parameters | Reference |
|---|---|---|
| Ellipse | Radii: 24.5 μm and 16.5 μm; Viscosity: draw from uniform distribution of values between 4 and 5 Pa.s. | Daniels *et al.*, 2006 |
| 2 solids (that mimic anterior and posterior centrosomes) | External force: Anterior: 180 pN; posterior: 500 pN. | Grill *et al.*, 2003 |
| 2 asters with two fibre types | Initial position: Anterior: -5.6 μm; Posterior 4.7 μm from ellipse centre. | *In vivo* lab measurements |
| Fibre, type #1 (spindle) | Activity: classic; Number per aster: 20; Initial length 5.4 μm (for 10) and 6.8 μm (for 10); Rigidity: 50 pN.μm$^2$; Position: 60° fan distribution. | |
| Fibre, type #2 (astral) | Activity: dynamic; Number per aster: 65 to 85 (random choice); Initial length $8 \pm 6$ μm; Rigidity: draw from Gaussian distribution of mean equal 40 pN.μm$^2$ and variance equal 5 pN.μm$^2$; Position: 240° aleatory fan distribution; Growing force: 5 pN; Minimal length: 0.005 μm; Growing speed: 0.71 μm/s; Shrinking sped: -0.84 μm/s; Catastrophe rate: 0.05 (no force); 0.5 (stall force); Rescue rate: 0.15. | Dogterom *et al.*, 1997 ; Srayko *et al.*, 2005 |
| Single with hands (that mimic cortical force generators) | Activity: bind; Anchored to a fixed position; Unbinding rate: 0.1; Unbinding force: 5 pN. | 99.4% |

In the second step, we generated synthetic images from these simulations using *ConfocalGN*, an image generator that mimics confocal microscopy [3]. To closely match our real images, we adopted an empirical approach rather than modelling fluorescence intensity analytically. Specifically, we extracted fluorescence intensity distributions for both microtubule and background pixels from real deconvolved images of live dividing embryos acquired using an Airyscan microscope (Fig. S1B, D2, E2). Synthetic fluorescent images were then constructed in two stages: (1) by applying the Point-Spread-Function (PSF) blur and simulating photon noise, and (2) by adding background noise (Fig. S1C). This process resulted in what we refer to as the "*MicSim_FluoMT simple* dataset". To better replicate uneven fluorescence observed along astral microtubules, we created a second dataset in which fluorescence intensity gradually decreases toward the cell periphery. We refer to as the "*MicSim_FluoMT complex* dataset", which introduced greater segmentation difficulty, providing a more rigorous test for segmentation algorithms. The images closely resembled real microscopy data, accurately capturing microtubule fluorescence intensity (Fig. S1E1), background noise levels (Fig. S1D1) and filament morphology (Fig. S1C3). Finally, we created binary masks to precisely annotate microtubules in each image. These masks were automatically derived from the simulation data, ensuring perfect alignment between the synthetic microtubule images and their corresponding annotations. They served as ground truth for training supervised deep learning models and for evaluating segmentation performance.

*A2. Synthetic images: characteristics, visualisation and advantages*

We conducted 500 simulation runs and selected several time points from each to construct the two *MicSim_FluoMT* datasets, each consisting of 1192 synthetic 2D images. The diversity in fluorescence intensity, microtubule density, and background noise highlights the dataset's complexity and its usefulness for training and evaluating deep learning models for microtubule segmentation. Fig. S2 provides an overview of the *MicSim_FluoMT* datasets, showcasing representative examples of the synthetic images alongside their corresponding annotations. Both datasets are publicly available on Zenodo (DOI 10.5281/zenodo.14696279) [4].

Existing datasets of fluorescently labelled filaments are primarily based on real confocal microscopy images, which often lack annotations or include only partial annotations. In addition, the limited number of available images may not fully capture the morphological diversity present in biological samples. In contrast, our synthetic fluorescent microscopy datasets offer several distinct advantages that make them excellent resources for filament segmentation tasks in microscopy. (1) Controlled variability: Unlike real datasets, our synthetic data allows full control over key image parameters such as noise levels, fluorescence intensity, and filament morphology. This enables systematic evaluation of model performance under various controlled conditions. (2) Broad range of conditions: The dataset includes both simple cases (e.g., isolated microtubules in low-noise environment) and more complex scenarios (e.g., overlapping filaments with high noise levels), providing comprehensive test conditions for segmentation models. (3) Comprehensive annotations: Each image in the dataset is fully annotated at the pixel level – something rarely available in real datasets – allowing for precise training and robust evaluation of deep learning models. (4) Large dataset size: Our dataset contains a large number of images, ensuring that deep learning models can be trained effectively and tested across diverse conditions, helping to improve their generalization and robustness.

B- Generation of a real dataset of stained microtubules, called *MicReal_FluoMT*

To evaluate the performance of our new architecture in segmenting microtubules in real images, we created a dataset comprising images of microtubules stained with an anti-tubulin antibody (DM1A-AF488 conjugate) in *Caenorhabditis elegans* zygotes. To ensure variability in microtubule shapes and densities, which is important for effective model training, we included four experimental conditions: two targeting *zyg-8*$^{\mathrm{DCLK1}}$ – a protein that binds to microtubules and regulates their rigidity – namely, *zyg-8(RNAi)* treated embryos and *zyg-8(or484ts)* heat-shocked mutants; and two control conditions, including RNAi control embryos and untreated heat-shocked embryos. We imaged fixed embryos using a confocal super-resolution microscope (Airyscan LSM980) and we acquired a stack of z-sections for each embryo. From each stack, we selected one or two z-sections that provided a clear visualisation of the astral microtubule network. When we selected two z-sections per image stack, we ensured they depicted distinct regions of the astral network. In total, we collected 49 images of *C. elegans* embryos, capturing two different stages of mitosis: metaphase and anaphase.

To annotate the microscopic images, we applied a three-step image processing pipeline, similar to the method described in [5]. (1) We performed an extended depth-of-field projection across 3 z-sections —specifically the section of interest along with the sections immediately above and below— to enhance microtubule continuity [6]. (2) We applied the Orientation Filter Transform to enhance filamentous pattern against noises [7]. (3) We applied the interactive machine-learning tool, *Ilastik,* to segment the astral microtubules [8]. During this semi-supervised segmentation, we manually annotated approximately 10 to 20 regions, labelling both microtubules and background in two to three embryos per condition. These annotations captured a range of intensities, contrasts, and microtubule morphologies, enabling the training of a segmentation model. This model was then applied to the remaining microscopy images to segment the astral microtubules in *C. elegans* embryos.

As a result, we obtained a dataset of 49 paired images, each consisting of a real microscopic image and a corresponding segmentation mask of the microtubules. We named this dataset "*MicReal_FluoMT*" and released it publicly on Zenodo (DOI 10.5281/zenodo.15852661) [9]. Of these, 19 images were used for training, 10 for validation, and 10 for testing. It is important to note that the semi-supervised segmentation occasionally identified structures outside the embryo that were not relevant to this study; however, these were retained in the segmented masks. Additionally, due to non-specific staining, the embryo periphery was often included in the segmentation masks. Some masks also contained small annotations within the embryo cytoplasm.

## 2- Supplemental figures

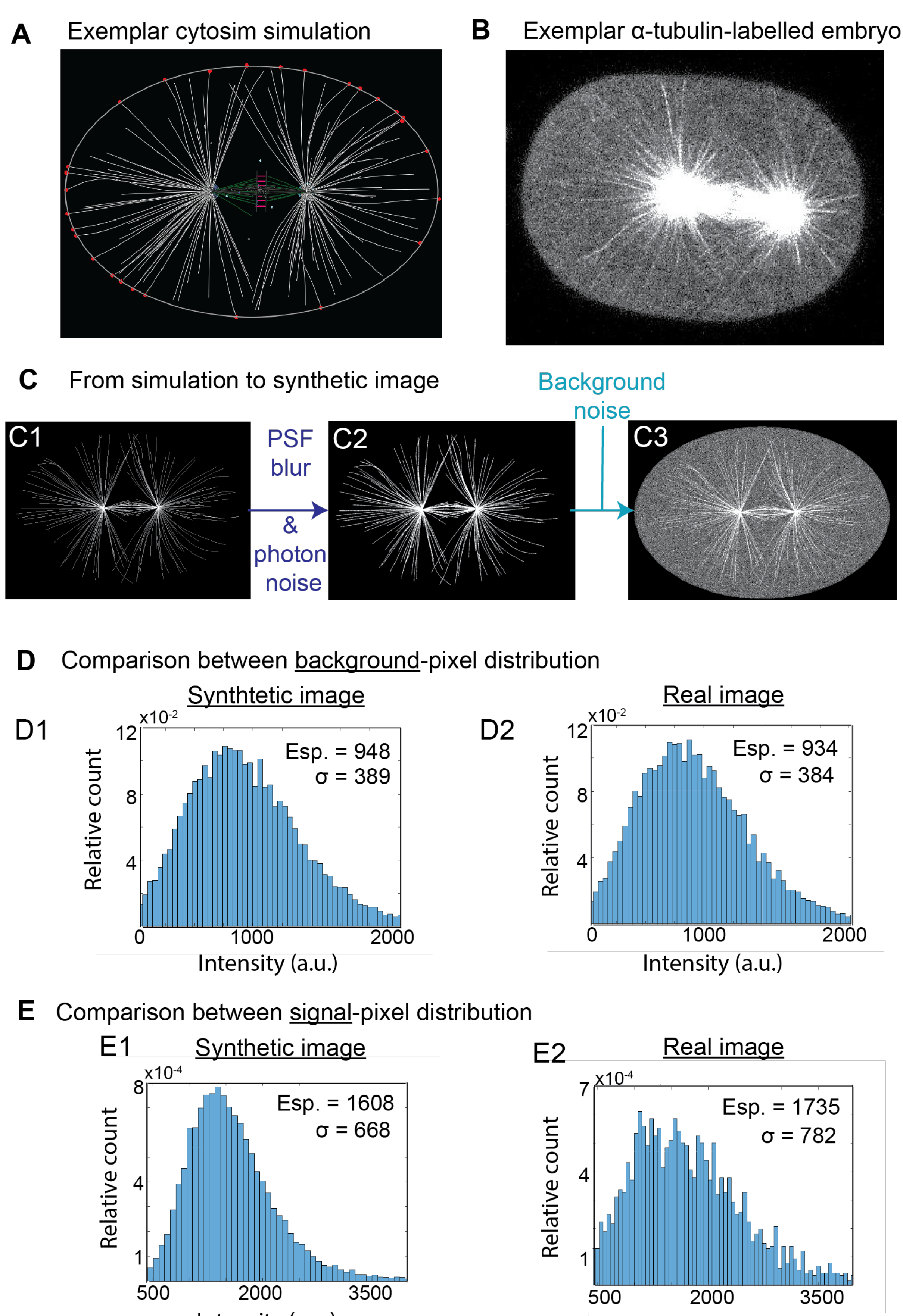

**Fig. S1: Generation of synthetic fluorescent images of microtubules for *MicSim_FluoMT* dataset, mimicking microtubule networks of *C. elegans* zygote visualised with fluorescently tagged tubulin.**

**(A)** Example of a *Cytosim* simulation showing astral microtubules (white lines); **(B)** Exemplar image of fluorescently labelled microtubules with YFP::$\alpha$-tubulin; **(C)** Two-stage generation of a synthetic image using *ConfocalGN*; **(D)** Comparison of background pixel intensity distributions between (D1) synthetic and (D2) real images; and **(E)** Comparison of microtubule pixel intensity distributions between (E1) synthetic and (E2) real images.

## Representatitve images of MicSim FluoMT easy and complex datasets

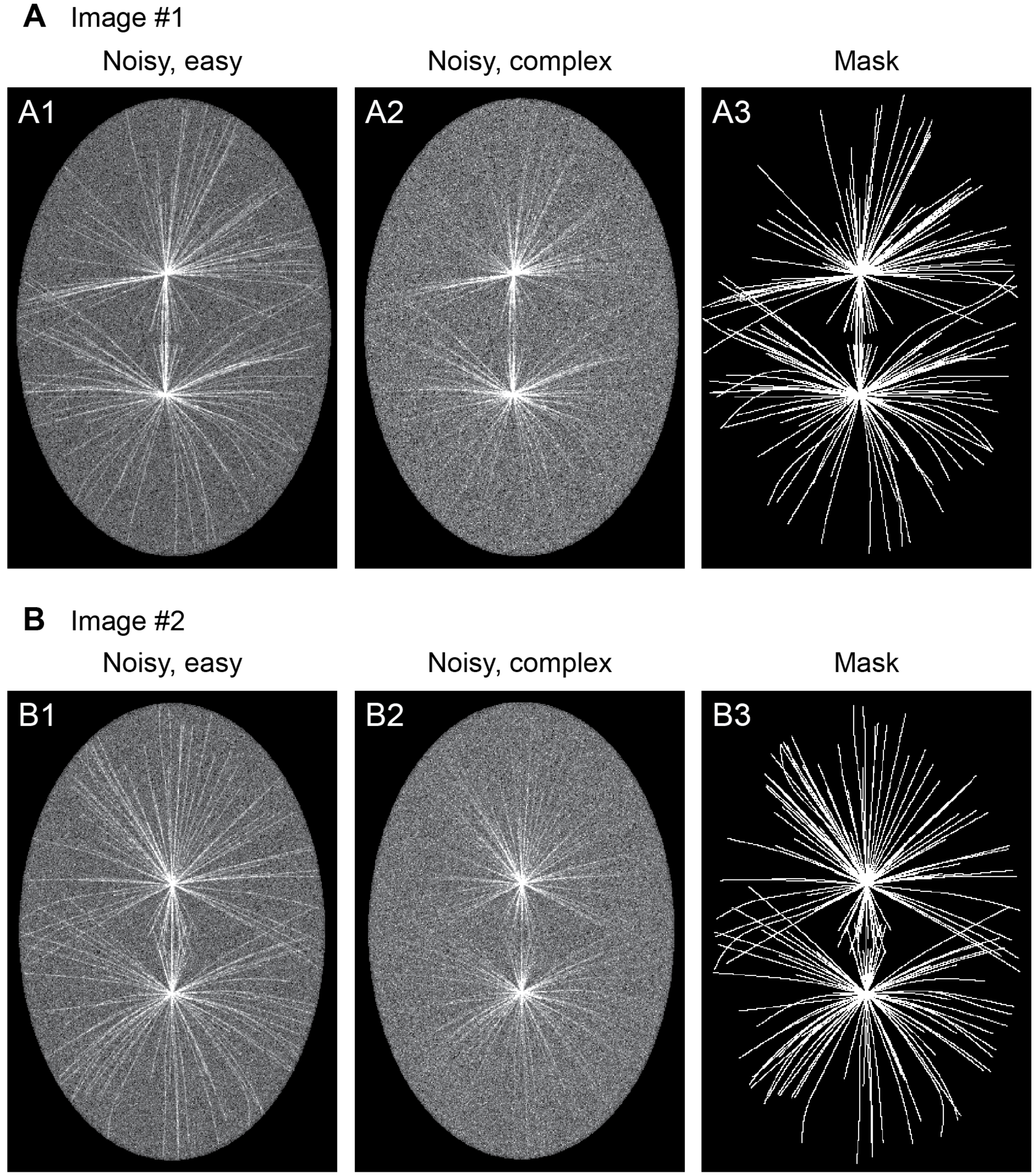

**Fig. S2: Overview of the two synthetic image datasets: *MicSim_FluoMT easy* and *MicSim_FluoMT complex*.**

Example images from *MicSim_FluoMT* datasets, with surrounding dark areas cropped to enhance filament visibility: (**A1, B1**) *Easy* images with uniform fluorescence along filaments and a signal-to-noise ratio (SNR) of (A1) 13.1 dB and (B1) 12.6 dB; the estimated noise levels *N*, calculated via 2D convolution (Eq. 2), are (A1) 0.336 and (B1) 0.346;(**A2, B2**) *Complex* images with uneven fluorescence and a SNR of (A2) 10.3 dB and (B2) 10.0 dB; the estimated noise levels *N* are (A2) 0.888 and (B2) 0.882; (**A3, C3**) Corresponding ground truth masks.

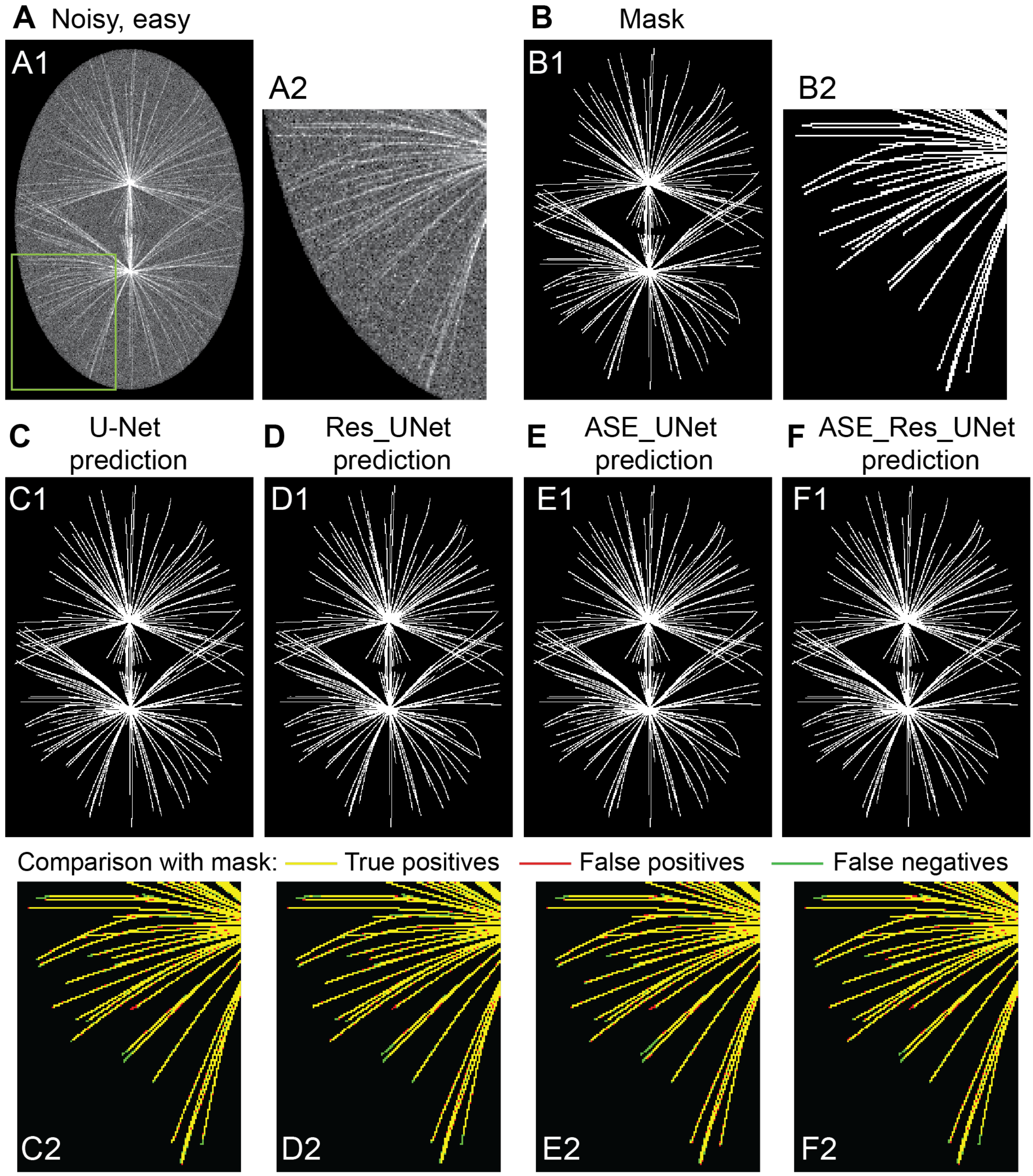

**Fig. S3: ASE_Res_UNet and its variants accurately segment microtubules in the *MicSim_FluoMT easy* dataset.**

Microtubule segmentation results obtained using ASE_Res_UNet and its variants on a sample test image from the *MicSim_FluoMT easy* dataset. (**A**) Input image; (**B**) Corresponding ground truth; (**C-F**) Predicted segmentations from (C) U-Net, (D) Res_UNet, (E) ASE_UNet, and (F) ASE_Res_UNet models. (A1-F1): whole simulated embryo; (A2-F2): zoomed-in regions of interest (ROI) to better highlight differences between model predictions and ground truth, with cropping window shown in green on panel A1; (C2-F2) composite ROI images showing true positives in yellow, false negatives in green, false positives in red, and true negatives in black.

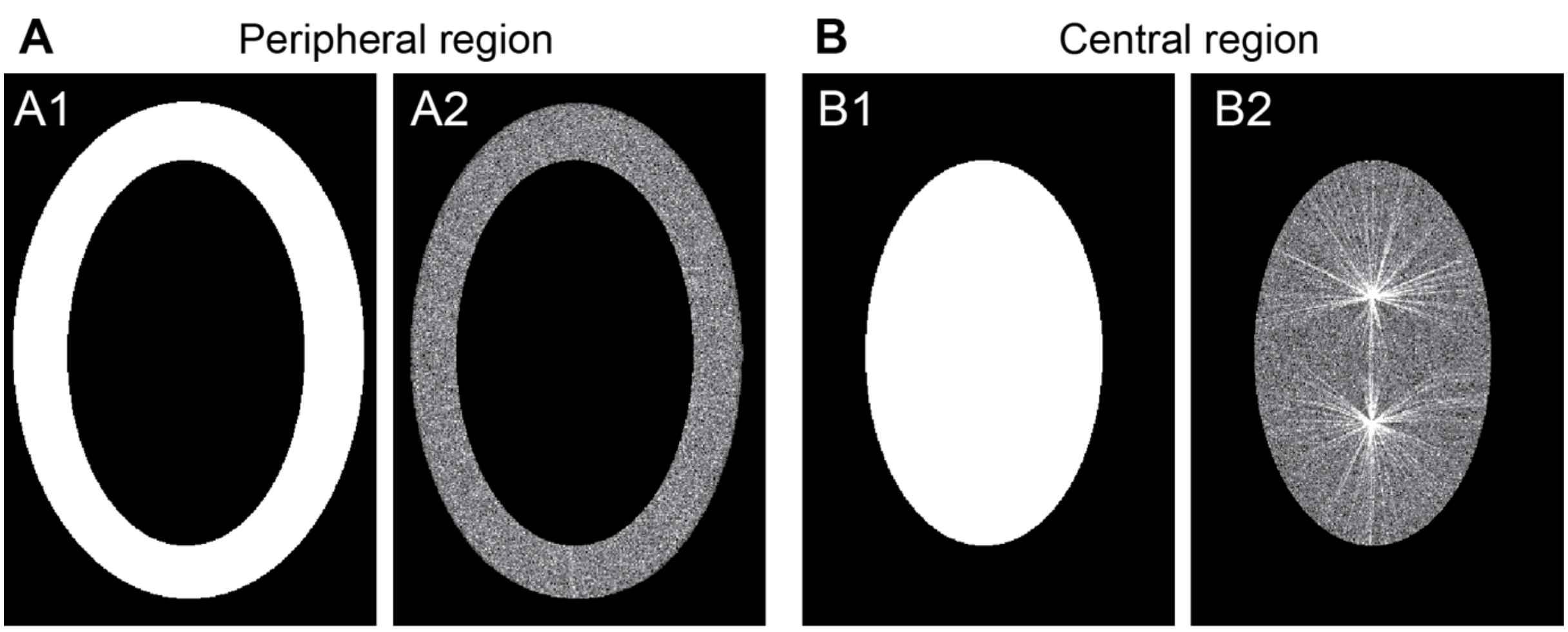

**Fig. S4: Definition of two embryo regions for spatial performance analysis in microtubule segmentation.**

Two regions of interest (ROI) were defined for spatial evaluation on the *MicSim_FluoMT complex* dataset: (**A**) peripheral region where microtubules have fluorescence intensities close to background noise; and (**B**) central region, where microtubules are more easily visible. (A1, B1) the masks of the ROI; and (A2, B2) ROIs overlaid on a representative test image.

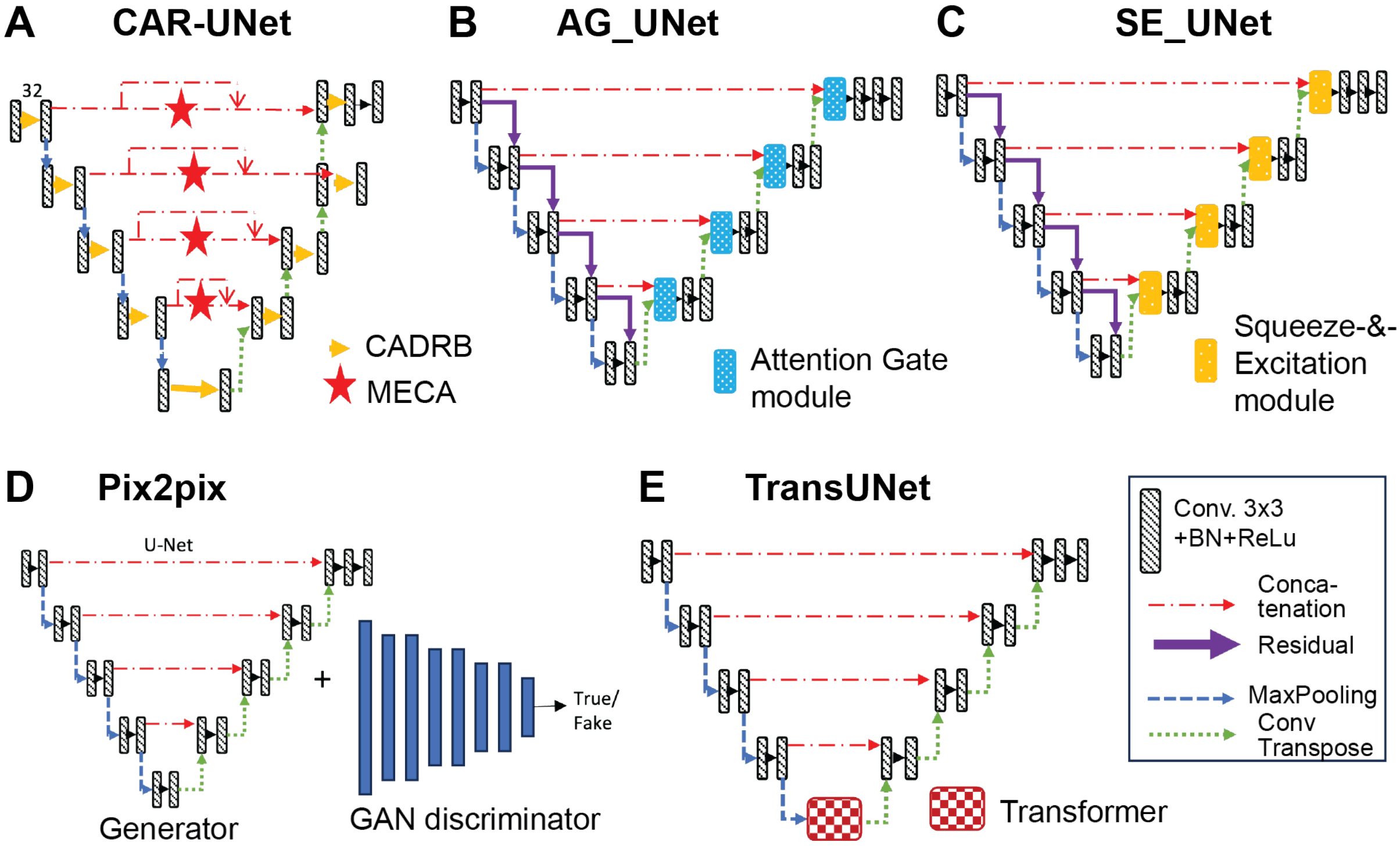

**Fig. S5: The five state-of-the-art architectures used for performance benchmarking in microtubule and vessel segmentation.**

Schematics of the architectures used to benchmark ASE_Res_UNet: (**A**) CAR-UNet, (**B**) AG_Res_UNet, (**C**) SE_Res_UNet, (**D**) Pix2pix, and (**E**) TransUNet.

## 3- Supplemental tables

| Metric | Description | Mathematical Formula | Relevance |
|---|---|---|---|
| **Dice** | Measures the overlap between predicted and ground truth masks. | $\frac{2TP}{2TP + FP + FN}$ | Good at evaluating segmentation where the positive class (e.g., filaments) is underrepresented; sensitive to boundary errors. |
| **IoU (Jaccard)** | Intersection over union between predicted and true regions. | $\frac{TP}{TP + FP + FN}$ | Stricter than Dice by penalizing more FP and FN; spatially accurate for verifying segmentation quality. |
| **Sensitivity** | Ability to correctly identify positive (foreground) pixels. | $\frac{TP}{TP + FN}$ | Crucial to avoid false negatives (i.e., missed curvilinear structures); ensures actual structures are captured. |
| **Precision** | Fraction of predicted positive pixels that are correct. | $\frac{TP}{TP + FP}$ | Important in noisy data; ensures that detected structures are truly relevant. |
| **MCC** | Correlation between predictions and true labels across all classes. | $\frac{TP \cdot TN - FP \cdot FN}{\sqrt{(TP + FP)(TP + FN)(TN + FP)(TN}}$ | Strong measure under class imbalance; reflects overall prediction quality. |
| **PR AUC** | Area under the precision-recall curve. | $\int_0^1 \text{Precision(Recall)}\ d(\text{Recall})$ | More suitable than ROC AUC for imbalanced data; focuses on relevance of predicted positive structures. |

**Table S1**: **Six selected metrics for assessing segmentation performance in the presence of background class dominance**.

The table provides the description, formula and relevance of the following metrics: the Dice coefficient, Intersection Over Union (IoU), Sensitivity, Precision, Matthews Correlation Coefficient (MCC), and Area Under the Precision-Recall Curve (PR AUC). TP, FP, FN, and TN denote True Positives, False Positives, False Negatives, and True Negatives, respectively.

| **Loss Function** | **Formula** | **Description** | **Relevance** |
|---|---|---|---|
| **Binary Cross Entropy (BCE)** | $L_{\text{BCE}} = -\frac{1}{N}\sum_{i=1}^{N}[y_i \log(\hat{y}_i) + (1-y_i)\log(1-\hat{y}_i)]$ | Standard loss for binary classification, treats all pixels equally. | Works well on balanced datasets but struggles with background-foreground imbalance typical in filamentous structures. |
| **Weighted Cross Entropy (WCE)** | $L_{\text{WCE}} = -\frac{1}{N}\sum_{i=1}^{N}\left[\begin{matrix} w_1 \times y_i \log(\hat{y}_i) \\ +w_0 \times (1-y_i)\log(1-\hat{y}_i) \end{matrix}\right]$ | Extension of BCE, assigns higher weight to underrepresented class (e.g., filaments). | Helps mitigate foreground-background imbalance, shown effective in noisy microscopy settings. |
| **Focal Loss** | $L_{\text{Focal}} = -\alpha(1-\hat{y}_i)^{\gamma} y_i \log(\hat{y}_i)$ With $\alpha = 0.25$ and $\gamma = 2$. | Focuses training on hard-to-classify examples by down-weighting easy ones. | Useful for datasets with extreme class imbalance or fine structures; however, tuning $\gamma$ and $\alpha$ is critical and does not promote structural accuracy. |
| **Dice Loss** | $L_{\text{Dice}} = 1 - \frac{2\sum y_i \hat{y}_i + \epsilon}{\sum y_i + \sum \hat{y}_i + \epsilon}$ | Directly optimizes for overlap between prediction and ground truth. | Helps preserve thin structures; however, may not capture fine topology and object boundary. |
| **Hausdorff Distance Loss** | Based on the Hausdorff distance: $H(A,B) = \max\{\sup_{a\in A}\inf_{b\in B} d(a,b), \sup_{b\in B}\inf_{a\in A} d(a$ | Penalizes segmentation errors based on spatial boundary discrepancies. | Strongly penalizes boundary mismatches, useful for assessing geometric accuracy of curvilinear shapes; however, limitations in case of class imbalance. |

**Table S2: Loss functions studied with their formula, description and relevance.**

$y_i$ is the ground truth label for the i-th pixel (0 for background, 1 for foreground) and $\hat{y}_i$ the predicted probability for the positive class (foreground) for the i-th pixel. N is the total number of pixels in the image. For WCE, $w_1$ and $w_0$ are the weights assigned to positive and negative classes, respectively. For the Focal loss, γ is a focusing parameter and α balances the importance of positive/negative examples. For Hausdorff Distance loss, *A* and *B* are sets of points on the contours of prediction and ground truth.

| Model | True Positive ↑ | True Negative ↑ | False Negative ↓ | False Positive ↓ |
|---|---|---|---|---|
| U-Net | 1 281 284 | 51 031 180 | 792 667 | 121 589 |
| Res_UNet | 1 306 883 | 50 998 731 | 767 068 | 154 038 |
| ASE_UNet | 1 455 872 | 50 902 052 | 618 079 | 250 717 |
| ASE_Res_UNet | 1 525 687 | 50 877 519 | 548 264 | 275 250 |

**Table S3: ASE_Res_UNet yields the highest number of True Positives and the lowest number of False Negatives, while maintaining a low number of False Positives.**

Confusion matrices of ASE_Res_UNet and its variant architectures, computed on the *MicSim_FluoMT complex* dataset (cumulative pixel count across 120 test images).

| Region | Model | Dice ↑ | IoU ↑ | Sensitivity ↑ | Precision ↑ | MCC ↑ | PR AUC ↑ |
|---|---|---|---|---|---|---|---|
| Central | U-Net | 0.8110 ± 0.0175 *** | 0.9797 ± 0.0016 *** | 0.7250 ± 0.0308 *** | **0.9216 ± 0.0142** *** | 0.8123 ± 0.0157 *** | 0.9256 ± 0.0095 * |
| | Res_UNet | 0.8117 ± 0.0164 *** | 0.9794 ± 0.0012 *** | 0.7365 ± 0.0269 *** | 0.9050 ± 0.0127 *** | 0.8112 ± 0.0153 *** | 0.9188 ± 0.0113 *** |
| | ASE_UNet | 0.8378 ± 0.0133 *** | 0.9812 ± 0.0011 *** | 0.8072 ± 0.0347 *** | 0.8736 ± 0.0344 ** | 0.8343 ± 0.0130 ** | 0.9271 ± 0.0106 |
| | ASE_Res_UNet | **0.8453 ± 0.0136** | **0.9817 ± 0.0009** | **0.8305 ± 0.0204** | 0.8608 ± 0.0098 | **0.8407 ± 0.0135** | **0.9297 ± 0.0101** |
| Peripheral | U-Net | 0.3595 ± 0.0526 *** | 0.9867 ± 0.0013 *** | 0.2315 ± 0.0427 *** | **0.8245 ± 0.0347 ***** | 0.4330 ± 0.0431 *** | 0.5605 ± 0.0534 ◊ |
| | Res_UNet | 0.3703 ± 0.0575 *** | 0.9866 ± 0.0012 *** | 0.2439 ± 0.0480 *** | 0.7882 ± 0.0360 *** | 0.4344 ± 0.0476 *** | 0.5391 ± 0.0556 *** |
| | ASE_UNet | 0.4448 ± 0.0541 *** | 0.9868 ± 0.0011 * | 0.3282 ± 0.0575 *** | 0.7115 ± 0.0659 * | 0.4778 ± 0.0454 *** | 0.5378 ± 0.0594 *** |
| | ASE_Res_UNet | **0.4997 ± 0.0488** | **0.9873 ± 0.0010** | **0.3933 ± 0.0520** | 0.6909 ± 0.0368 | **0.5175 ± 0.0433** | **0.5770 ± 0.0534** |

**Table S4: The performance gain of ASE_Res_UNet over its architectural variants is more pronounced in the peripheral region.**

Microtubule segmentation performances of ASE_Res_UNet and its variants on the *MicSim_FluoMT complex* dataset, evaluated in the peripheral and central regions of the embryo using various metrics (mean ± standard deviation over 120 test images). Bold values indicate the best performances for each metric. Statistical differences between ASE_Res_UNet and its variants are indicated only when significant (◊: $0.01 < p \leq 0.05$; *: $0.001 < p \leq 0.01$; **: $0.0001 < p \leq 0.001$; ***: $p \leq 0.0001$).

| Model | True Positive ↑ | True negative ↑ | False negative ↓ | False positive ↓ |
|---|---|---|---|---|
| CAR-UNet | 1 398 010 | 50 952 673 | 675 941 | 200 096 |
| AG_Res_UNet | 1 404 812 | 50 954 420 | 669 139 | 198 349 |
| SE_Res_UNet | 1 418 884 | 50 944 778 | 655 067 | 207 991 |
| TransUNet | 1 499 852 | 50 848 011 | 574 099 | 304 758 |
| Pix2pix | 1 412 662 | 50 543 648 | 661 289 | 609 121 |
| ASE_Res_UNet | 1 526 218 | 50 857 777 | 547 733 | 294 992 |

**Table S5: Compared to state-of-the-art models, ASE_Res_UNet yields the highest number of True Positives and the lowest number of False Negatives, while maintaining a low number of False Positives.**

Confusion matrices of ASE_Res_UNet and other advanced architectures, computed on the *MicSim_FluoMT complex* dataset (cumulative pixel count across 120 test images).

| Region | Model | Dice ↑ | IoU ↑ | Sensitivity ↑ | Precision ↑ | MCC ↑ | PR AUC ↑ |
|---|---|---|---|---|---|---|---|
| Central | CAR-UNet | 0.8309 ± 0.0137 *** | 0.9809 ± 0.0010 *** | 0.7802 ± 0.0207 *** | 0.8890 ± 0.0135 *** | 0.8279 ± 0.0134 *** | 0.9242 ± 0.0109 *** |
| | AG_Res_UNet | 0.8318 ± 0.0147 *** | 0.9810 ± 0.0010 *** | 0.7799 ± 0.0201 *** | **0.8913 ± 0.0137** *** | 0.8269 ± 0.0145 *** | 0.9265 ± 0.0115 ◊ |
| | SE_Res_UNet | 0.8338 ± 0.0160 *** | 0.9812 ± 0.0010 *** | 0.7858 ± 0.0241 *** | 0.8885 ± 0.0110 *** | 0.8307 ± 0.0159 *** | 0.9263 ± 0.0116 ◊ |
| | Pix2pix | 0.7549 ± 0.0166 *** | 0.9700 ± 0.0011 *** | 0.7712 ± 0.0141 *** | 0.7396 ± 0.0252 *** | 0.7473 ± 0.0166 *** | 0.8193 ± 0.0196 *** |
| | TransUNet | 0.8368 ± 0.0151 *** | 0.9808 ± 0.0009 *** | 0.8169 ± 0.0240 *** | 0.8579 ± 0.0084 ◊ | 0.8321 ± 0.0149 *** | 0.9221 ± 0.0118 *** |
| | ASE_Res_UNet | **0.8453 ± 0.0136** | **0.9817 ± 0.0009** | **0.8305 ± 0.0204** | 0.8608 ± 0.0098 | **0.8407 ± 0.0135** | **0.9297 ± 0.0101** |
| Peripheral | CAR-UNet | 0.4168 ± 0.0503 *** | 0.9869 ± 0.0011 * | 0.2905 ± 0.0460 *** | 0.7497 ± 0.0386 *** | 0.4628 ± 0.0423 *** | 0.5471 ± 0.0525 *** |
| | AG_Res_UNet | 0.4336 ± 0.0480 *** | 0.9870 ± 0.0011 | 0.3068 ± 0.0456 *** | **0.7538 ± 0.0565 *** ** | 0.4768 ± 0.0414 *** | 0.5663 ± 0.0551 |
| | SE_Res_UNet | 0.4386 ± 0.0540 *** | 0.9870 ± 0.0011 | 0.3136 ± 0.0519 *** | 0.7409 ± 0.0343 *** | 0.4781 ± 0.0457 *** | 0.5530 ± 0.0568 ** |
| | Pix2pix | 0.4092 ± 0.0390 *** | 0.9833 ± 0.0012 *** | 0.3580 ± 0.0418 *** | 0.4807 ± 0.0464 *** | 0.4101 ± 0.0384 *** | 0.4069 ± 0.0496 *** |
| | TransUNet | 0.4824 ± 0.0536 * | 0.9868 ± 0.0011 ** | 0.3816 ± 0.0589 | 0.6629 ± 0.0325 *** | 0.4989 ± 0.0472 * | 0.5450 ± 0.0574 *** |
| | ASE_Res_UNet | **0.4997 ± 0.0488** | **0.9873 ± 0.0010** | **0.3933 ± 0.0520** | 0.6909 ± 0.0368 | **0.5175 ± 0.0433** | **0.5770 ± 0.0534** |

**Table S6: ASE_Res_UNet outperforms state-of-the-art models in both central and peripheral regions, with the largest performance margin observed in the periphery.**

Microtubule segmentation performances of ASE_Res_UNet and other advanced architectures on the *MicSim_FluoMT complex* dataset, evaluated separately in embryo peripheral and central regions using various metrics (mean ± standard deviation over 120 test images). Bold values indicate the best performances for each metric. Statistical differences between ASE_Res_UNet and other architectures are indicated only when significant (◊: $0.01 < p \leq 0.05$; *: $0.001 < p \leq 0.01$; **: $0.0001 < p \leq 0.001$; ***: $p \leq 0.0001$).

## 4- References